\documentclass[cpp,a4paper,fleqn%
]{w-art}
\usepackage{times,cite,w-thm}

\theoremstyle{plain}

\theoremstyle{definition}

\usepackage[]{graphicx}
\begin{document}
\DOIsuffix{theDOIsuffix}
\Volume{xx}
\Month{xx}
\Year{xxxx}
\pagespan{1}{}
\Receiveddate{XXXX}
\Reviseddate{XXXX}
\Accepteddate{XXXX}
\Dateposted{XXXX}



\title[Factorization of structure functions in Yukawa liquids]{Factorization of 3-point static structure functions in 3D Yukawa liquids}


\author[P. Magyar]{P\'eter Magyar\inst{1}\footnote{Corresponding author \quad E-mail:~\textsf{magyar.peter@wigner.mta.hu}}}
\address[\inst{1}]{Institute for Solid State Physics, Wigner Research Centre for Physics, Hungarian Academy of Sciences, H-1525 Budapest, P.O. Box 49, Hungary}
\author[P. Hartmann]{Peter Hartmann\inst{1,2}}
\author[G. J. Kalman]{Gabor J. Kalman\inst{2}}
\address[\inst{2}]{Department of Physics, Boston College, Chestnut Hill, Massachusetts 02467, USA}
\author[K. I. Golden]{Kenneth I. Golden\inst{3}}
\address[\inst{3}]{Department of Mathematics and Statistics, and Department of Physics, University of Vermont, Burlington, USA}
\author[Z. Donk\'o]{Zolt\'an Donk\'o\inst{1,2}}

\begin{abstract}
In many-body systems the convolution approximation states that the 3-point static structure function, $S^{(3)}(\textbf{k}_{1},\textbf{k}_{2})$, can approximately be ``factorized'' in terms of the 2-point counterpart, $S^{(2)}(\textbf{k}_{1})$. We investigate the validity of this approximation in 3-dimensional strongly-coupled Yukawa liquids: the factorization is tested for specific arrangements of the wave vectors $\textbf{k}_{1}$ and $\textbf{k}_{2}$, with molecular dynamics simulations. With the increase of the coupling parameter we find a breakdown of factorization, of which a notable example is the appearance of negative values of $S^{(3)}(\textbf{k}_{1},\textbf{k}_{2})$, whereas the approximate factorized form is restricted to positive values. These negative values -- based on the quadratic Fluctuation-Dissipation Theorem -- imply that the quadratic part of the density response of the system changes sign with wave number. Our simulations that incorporate an external potential energy perturbation clearly confirm this behavior.
\end{abstract}

\maketitle                   

\section{Introduction}

The dynamical and static structure functions are of paramount importance in describing the dynamics and the structure of many-body systems. These quantities can be measured in scattering experiments (involving electrons, neutrons and x-ray radiation) on substances like liquids \cite{White,Pothoczki,Hura}, glasses \cite{Hirata,Ma} and dense plasmas \cite{Thiele1,Thiele2,Murillo1,Souza,Riley}.

The conventional dynamical structure function $S^{(2)}({\bf k}_{1},\omega_{1})$, is defined as the Fourier transform of the 2-point dynamical density correlation
\begin{equation}
S^{(2)}({\bf k}_{1},\omega_{1})=\frac{1}{N}\int_{-\infty}^{\infty}e^{i\omega_{1} t}\langle n(\textbf{k}_{1},t) ~n(-\textbf{k}_{1},0)\rangle^{(0)} {\rm d}t,
\end{equation}
whereas its static (equal-time) counterpart is given by
\begin{equation}
S^{(2)}({\bf k_{1}})=\frac{1}{N}\langle n(\textbf{k}_{1},t) ~n(-\textbf{k}_{1},t)\rangle^{(0)}.
\label{linsk}
\end{equation}

It is not difficult to realize, though, that the 2-point structure function is in fact the lowest rank member of a hierarchy of more general $p$-point  structure functions, the Fourier transforms of $p$-point density correlations \cite{Golden-Heath}:
\begin{eqnarray}
S^{(p+1)}(\textbf{k}_{1},\textbf{k}_{2},...,\textbf{k}_{p};\omega_{1},\omega_{2},...,\omega_{p})\delta_{\textbf{k}_{1}+\textbf{k}_{2}+...+\textbf{k}_{p}+\textbf{k}_0,0}
\delta(\omega_{1}+\omega_{2}+...+\omega_{p}+\omega)=& \\
\frac{1}{N}\int_{-\infty}^{\infty}{\rm d}t_{1}e^{i\omega_{1}t_{1}}\int_{-\infty}^{\infty}{\rm d}t_{2}e^{i\omega_{2}t_{2}}...\int_{-\infty}^{\infty}{\rm d}t_{p}e^{i\omega_{p}t_{p}}
\langle n(\textbf{k}_{1},t_{1})~n(\textbf{k}_{2},t_{2})...n(\textbf{k}_{p},t_{p})~n(\textbf{k}_0,0)\rangle^{(0)}=& \nonumber \\
\frac{1}{2\pi N}\langle n(\textbf{k}_{1},\omega_{1})~n(\textbf{k}_{2},\omega_{2})...n(\textbf{k}_{p},\omega_{p})~n(\textbf{k}_0,\omega)\rangle^{(0)}. \nonumber
\end{eqnarray}
Note that we use a slightly changed notation here compared to that used in \cite{Golden-Heath}, to maintain the convention used in our previous work \cite{Magyar}. The higher rank structure functions have enjoyed less of a popularity, but it has been recognized that they may provide important additional information beyond $S^{(2)}(\textbf{k}_{1},\omega_{1})$, on the dynamics of many body systems \cite{Efremov,Bochkov,Evans}.

It is well-known that $S^{(2)}({\bf k}_{1},\omega_{1})$ is linked via the centrally important Fluctuation-Dissipation Theorem (FDT) \cite{Kubo} with the linear response of many-body systems. In a similar fashion, it is expected that the higher rank structure functions are linked via the higher-order FDT-s with the nonlinear response of these systems \cite{NFDT,Sitenko,BGY}. The explicit relationships have been recently worked out to the  fourth order, where the quartic FDT establishes a, rather complex, connection between quartic response functions and 5-point structure functions \cite{Golden-Heath}.

In the present paper we are concerned with the static (equal-time) limit of $S^{(3)}(\textbf{k}_{1},\textbf{k}_{2},\omega_{1},\omega_{2})$,
\begin{equation}
S^{(3)}(\textbf{k}_{1},\textbf{k}_{2})\delta_{\textbf{k}_{1}+\textbf{k}_{2}+\textbf{k}_0,0}=\frac{1}{N}\langle n(\textbf{k}_{1},t) ~n(\textbf{k}_{2},t) ~n(\textbf{k}_0,t)\rangle^{(0)}
\label{quadsk}
\end{equation}
and the corresponding static (zero-frequency) quadratic response function $\hat{\chi}^{(2)}(\textbf{k}_{1},\textbf{k}_{2})$.

There is also a well known intimate relationship between the 2-point static structure function and the equilibrium 2-particle distribution function $g^{(2)}({\bf r}_1,{\bf r}_2)$ with  its companion  2-particle correlation function $h^{(2)}({\bf r}_1,{\bf r}_2)$; similarly linked are the 3-point static structure function and the 3-particle distribution and correlation functions $g^{(3)}({\bf r}_1,{\bf r}_2,{\bf r}_3)$ and $h^{(3)}({\bf r}_1,{\bf r}_2,{\bf r}_3)$. To see this, we recall the conventionally defined 2-particle and 3-particle distribution functions, and define the respective pair and triplet correlation functions by
\begin{equation}
g^{(2)}({\bf r}_1,{\bf r}_2) \equiv g^{(2)}({\bf r}_{12}) = 1+h^{(2)}({\bf r}_{12})
\end{equation}
and
\begin{equation}
g^{(3)}({\bf r}_1,{\bf r}_2,{\bf r}_3) \equiv g^{(3)}({\bf r}_{12}, {\bf r}_{23}) =
1+h^{(2)}({\bf r}_{12})+h^{(2)}({\bf r}_{23})+h^{(2)}({\bf r}_{31})+h^{(3)}({\bf r}_{12}, {\bf r}_{23}).
\end{equation}	
The Fourier transforms of the correlation functions  are linked  to the respective structure functions by the linear
\begin{equation}
S^{(2)}({\bf k}_1) = 1+ n_{0} h^{(2)}({\bf k}_1)
\label{s1}
\end{equation}
and quadratic \cite{NFDT}
\begin{eqnarray}
S^{(3)}({\bf k}_1,{\bf k}_2) =
1+ n_{0} [ h^{(2)}({\bf k}_1)+ h^{(2)}({\bf k}_2)+ h^{(2)}({\bf k}_0)] + n_{0}^2 h^{(3)}({\bf k}_1,{\bf k}_2)& \\
\textbf{k}_{0}+\textbf{k}_{1}+\textbf{k}_{2}=0 \nonumber
\label{s2}
\end{eqnarray}
relationships, where $n_{0}$ is the macroscopic density.

The question whether the structure functions of higher ranks can be reduced (at least approximately) to appropriately defined clusters of structure functions of lower rank has been attracting considerable attention over the past decades. In particular, the possibility of the ``factorization'' of the 3-point structure function, $S^{(3)}({\bf k}_1,{\bf k}_2)$, in terms of 2-point structure functions, has been studied by several authors \cite{Jackson,O'Neil,Lie,Ichimaru,Hansen,Ichimaru2}.

The purpose of this work is to investigate the factorization properties of the 3-point structure functions by focusing on a particular system of great current interest, the (three-dimensional) Yukawa-liquid. In this system the potential between the particles has the form of an exponentially screened Coulomb potential,
\begin{equation}
\phi(r) = \frac{Q}{4 \pi \varepsilon_0} \frac{\exp(-r / \lambda_{\rm D})} {r},
\end{equation}
where $Q$ is the charge of the particles and $\lambda_{\rm D}$ is the screening (Debye) length. Apart from its applicability to actual physical systems, the Yukawa potential as a theoretical model has the convenient feature that by adopting limits of the screening parameter one can investigate both the properties of a system with  long range interaction $(\lambda_{\rm D}\rightarrow \infty)$ and short range interaction $(\lambda_{\rm D}\rightarrow 0)$.

The interest in Yukawa systems is also motivated by ongoing studies of dusty (complex) plasmas \cite{Complex1,Complex2,Complex3}. In the standard model for dusty plasmas one considers explicitly only one type of charged species (the dust particles), the presence of the additional plasma species and their effects on the main component are accounted for via the screening of the interaction potential between the dust particles. When all effects that lead to an anisotropy of the interaction (e.g. streaming ions) are neglected, the potential between the particles acquires the exponentially screened Coulomb form, becoming a Debye-H\"uckel, or Yukawa potential, where  the Debye length that expresses the screening property of the plasma medium that embeds the system of dust particles. (The system so defined is sometimes referred to as the ``Yukawa One-Component Plasma'', YOCP, based on its similarities to the classical Coulomb ``One-Component Plasma, OCP, model). The ratio of the unscreened inter-particle potential energy to the thermal energy is expressed by the nominal coupling parameter
\begin{equation}
\Gamma = \frac{Q^2}{4 \pi \varepsilon_0 a k_B T},
\end{equation}
where $T$ is temperature and $a = (3/4 \pi n_0)^{1/3}$ is the Wigner-Seitz radius. The normalization of the length by $a$ leads to the dimensionless screening parameter $\kappa = a / \lambda_{\rm D}$ . We investigate the system in the strongly coupled liquid phase ($\Gamma \gg 1$), where the interparticle potential energy dominates over the thermal energy and a prominent liquid structure builds up. It is not the subject of our studies, but we note that the system turns into a crystal when the coupling parameter reaches a certain critical value that depends on the screening parameter \cite{Hamaguchi}.

The structure of the paper is as follows. In Section 2 we introduce the quantities of interest and in Section 3 we describe our computational approach. Section 4 presents the results and a short summary is given in Section 5.

\section{Theoretical background}

Exact relationships between the 2- and 3-point static structure functions (or equivalently the respective particle correlation functions) exist. In particular, the second equation of the BGY hierarchy can be expressed in terms of the 2-point and 3-point static structure functions \cite{BGY}. However, this is not immediately useful, because it merely expresses one unknown function, the 2-point structure function (or 2-particle correlation function), in terms of another, 3-point structure function (or 3-particle correlation function), and it requires an approximate closure relation to become solvable. A further exact relationship is based on the density functional formalism \cite{Hansen}, where   in addition to the conventional (binary)  direct correlation function $c^{(2)}(\textbf{r}_{12})$ the
ternary direct correlation function $c^{(3)}(\textbf{r}_{12},\textbf{r}_{23})$ can be introduced. Then, \begin{equation}
S^{(3)}(\textbf{k}_{1},\textbf{k}_{2};\textbf{k}_{0})=S^{(2)}(\textbf{k}_{1})S^{(2)}(\textbf{k}_{2})S^{(2)}(\textbf{k}_{0})
\big[1+n_{0}^{2}c^{(3)}(\textbf{k}_{1},\textbf{k}_{2}) \big],
\label{OZ}
\end{equation}
where $c^{(3)}$ can be  obtained from the free energy functional \cite{Hansen}. This expression, however, does not provide any calculational guidance either without further approximation. Provided, however, that if $c^{(3)}(\textbf{k}_{1},\textbf{k}_{2})$ is small, the 3-point static structure function $S^{(3)}(\textbf{k}_{1},\textbf{k}_{2})$ can, at least approximately, be factorized in terms of 2-point static structure functions:
\begin{equation}
S^{(3)}({\bf k}_1,{\bf k}_2) \cong S^{(2)}({\bf k}_1) S^{(2)}({\bf k}_2) S^{(2)}({\bf k}_0).
\label{factorization}
\end{equation}

We note that the factorization (\ref{factorization}) leads to an approximate relation between the 2-particle and 3-particle correlation functions. In view of the relations (\ref{s1}) and (\ref{s2}) the relationship (\ref{factorization}) can be expressed  as
\begin{eqnarray}
h^{(3)}(\textbf{k}_1,\textbf{k}_2) \cong  h^{(2)}(\textbf{k}_1) h^{(2)}(\textbf{k}_2) + h^{(2)}(\textbf{k}_2) h^{(2)}(\textbf{k}_0)
+ h^{(2)}(\textbf{k}_0) h^{(2)}(\textbf{k}_1) + & \nonumber \\
n_{0} \big[ h^{(2)}(\textbf{k}_1) h^{(2)}(\textbf{k}_2) h^{(2)}(\textbf{k}_0) \big],
\label{hk1k2}
\end{eqnarray}
or in configuration space as
\begin{eqnarray}
h^{(3)}({\bf r}_{12},{\bf r}_{23}) \cong
n_{0} \Biggl[ \int {\rm d}^3 {\bf r}_4 h^{(2)}({\bf r}_{14})h^{(2)}({\bf r}_{42})+ \int {\rm d}^3 {\bf r}_4 h^{(2)}({\bf r}_{24})h^{(2)}({\bf r}_{43})+ & \nonumber \\
\int {\rm d}^3 {\bf r}_4 h^{(2)}({\bf r}_{34})h^{(2)}({\bf r}_{41})+
\int {\rm d}^3 {\bf r}_4 h^{(2)}({\bf r}_{14}) h^{(2)}({\bf r}_{24}) h^{(2)}({\bf r}_{34}) \Biggr].
\label{hrr}
\end{eqnarray}

The approximation (\ref{factorization}) was originally introduced in \cite{Jackson} and was called the ``convolution approximation''. O'Neil and Rostoker \cite{O'Neil}, Lie and Ichikawa \cite{Lie} showed that the expression (\ref{hk1k2}) is formally exact to lowest order of the plasma parameter and small $k$ values. Ichimaru \cite{Ichimaru} used the approximation (\ref{hrr}) as an ansatz to solve the second equation of the BBGKY hierarchy and so to calculate the dielectric response function for a strongly correlated electron liquid. Golden and Kalman \cite{NFDT} pointed out that in the framework of the quadratic FDT (\ref{hrr}) is a trivial  consequence of the application of the RPA and is valid for arbitrary $k$-values.

Remarkably though, as discussed by Iyetomi  and Ichimaru \cite{Ichimaru2}, while not exact, (\ref{hk1k2}) still represents a reasonably good approximation for the OCP up to a fairly large value of the coupling parameter $\Gamma$. In the language of the Mayer cluster expansion, the approximation amounts to the neglect of the irreducible bridge-functions, i.e. to invoking the Hypernetted Chain (HNC) approximation. This latter has been shown to provide a good description of the equilibrium properties of a strongly coupled OCP. Iyetomi  and Ichimaru \cite{Ichimaru2} also identified the group of  bridge functions that can be generated by a systematic expansion  in order to go beyond the HNC and have shown that the first correction to the convolution approximation is:
\begin{equation}
c^{(3)}({\bf r},{\bf r}') \approx h^{(2)}(r)  h^{(2)}(r) h^{(2)}(|{\bf r}-{\bf r}'|).
\label{eq:hhh}
\end{equation}
Based on this expression, Barrat, Hansen and Pastore \cite{Hansen} proposed the following factorization approximation for the triplet direct correlation function in dense, classical fluids (referred to as the ``BHP model'' in the following) :
\begin{equation}
c^{(3)}({\bf r},{\bf r}') = t(r)  t(r') t(|{\bf r}-{\bf r}'|),
\label{eq:bhp1}
\end{equation}
where the $t(r)$ function originates from the solution of the integral equation:
\begin{equation}
\partial c^{(2)}(r) \partial n = t(r) \int  t(r') t(|{\bf r}-{\bf r}'|) {\rm d} {\bf r'}.
\label{eq:bhp2}
\end{equation}
The predictions of this theory were tested by computing the 2-point and 3-point structure functions via molecular dynamics simulations for the soft sphere model near freezing and by using these in the relation (\ref{OZ}). The BHP model was successfully applied in, e.g., \cite{binary,qYukawa,water} for a binary mixture of hard spheres, quantum hard-sphere Yukawa fluids, and liquid water.

The purpose of this paper is to examine in some detail the accuracy of the approximation (\ref{factorization}) and the performance of the BHP model in Yukawa-liquids. The simplest configuration of the three $k$-vectors involved in the approximation is when all of them are collinear:  this scenario was studied in our previous paper \cite{Magyar}. Here we consider some special geometries when the vector-arguments of $S^{(3)}$ are not collinear.

\section{Computational method}

Molecular dynamics (MD) simulations have been indispensable in the studies of static properties, transport coefficients, collective excitations, as well as instabilities in strongly coupled Yukawa liquids. Here we use a standard MD method to describe our 3-dimensional Yukawa liquid: we simulate the motion of $N$=10~000 particles, within a cubic box with periodic boundary conditions, via the integration of their Newtonian equations of motion. The exponential decay of the Yukawa potential makes it possible to use a cutoff radius in the calculation of the forces acting on the particles, beyond which the interaction of particle pairs can be neglected. Setting $r_{\rm cutoff} \approx 5 a$ provides good accuracy at the relatively high value of the screening parameter considered in this work, $\kappa=3$. Time integration is performed via the velocity-Verlet scheme. At the initialization of the simulations the particles are positioned randomly within the simulation box, while their initial velocities are sampled from a Maxwellian distribution corresponding to the prescribed system temperature (defined by $\Gamma$). The simulations start with a thermalization phase, during which the particle velocities are rescaled in each time step, in order to reach the desired temperature. This procedure is stopped before the data collection takes place, where the stability of the simulation is confirmed by monitoring the temperature as a function of time.

The static structure functions $S^{(2)}$ and $S^{(3)}$ are computed in equilibrium MD simulations, via definitions (\ref{linsk}) and (\ref{quadsk}), for a series of wave number values matching the simulation box. The minimum accessible wave number defined by the edge length $L$ of the simulation box is $k_{min}a = 2 \pi a/L \cong 0.181$. The computations give direct access to wave number vectors that obey the periodic boundary conditions, i.e. to a ``grid'' of wave numbers with ${\bf k} = (m_x,m_y,m_z) k_{\rm min}$, where $m_x, m_y, m_z$ are integers (with the $m_x=0, m_y=0, m_z=0$ combination being excluded). In some of the cases the structure functions have to be computed for ${\bf k}$ values different from the above, in these circumstances we use an interpolation of the values from the four neighboring points that are on the grid.

To address the question: ``How accurate is the approximate factorization relation for our system?'', we calculate the following quantities, based on (\ref{s2}) and (\ref{hk1k2}):
\begin{eqnarray}
S^{(3)}_{f}({\bf k}_1,{\bf k}_2) & \equiv & S^{(2)}({\bf k}_1) S^{(2)}({\bf k}_2) S^{(2)}({\bf k}_0),
\label{Sf}\\
n_{0}^2 h^{(3)}_{m}({\bf k}_1,{\bf k}_2) & \equiv & S^{(3)}_{m}({\bf k}_1,{\bf k}_2)-1
-n_{0} [ h^{(2)}({\bf k}_1)+ h^{(2)}({\bf k}_2)+ h^{(2)}({\bf k}_0)],
\label{hm}\\
n_{0}^2 h^{(3)}_{f}(\textbf{k}_1,\textbf{k}_2) & \equiv & n_{0}^2\big[ h^{(2)}(\textbf{k}_1) h^{(2)}(\textbf{k}_2) + h^{(2)}(\textbf{k}_2) h^{(2)}(\textbf{k}_0)
+ h^{(2)}(\textbf{k}_0) h^{(2)}(\textbf{k}_1)\big] + \nonumber \\
&& n^{3}_{0} \big[ h^{(2)}(\textbf{k}_1) h^{(2)}(\textbf{k}_2) h^{(2)}(\textbf{k}_0) \big],
\label{hf}
\end{eqnarray}
where $S^{(3)}_{m}({\bf k}_1,{\bf k}_2)$ is the ``exact'' value of the 3-point structure function resulting from its measurement in the simulation, according to (\ref{quadsk}). $h^{(2)}$ is obtained from $S^{(2)}$ via (\ref{s1}). To describe the accuracy of the factorization approximation we compute the following quantities:
\begin{eqnarray}
\sigma_{S}(\textbf{k}_1,\textbf{k}_2)& \equiv & \frac{S^{(3)}_{m}(\textbf{k}_1,\textbf{k}_2)}{S^{(3)}_{f}(\textbf{k}_1,\textbf{k}_2)} \label{RS}, \\
\sigma_{h}(\textbf{k}_1,\textbf{k}_2)& \equiv & \frac{h^{(3)}_{m}(\textbf{k}_1,\textbf{k}_2)}{h^{(3)}_{f}(\textbf{k}_1,\textbf{k}_2)}.
\label{Rh}
\end{eqnarray}

Both of these give a relation between the ``exact'' characteristics and their approximate values obtained with the assumption that the direct factorization relation (\ref{factorization}) holds, i.e. $c^{(3)}=0$; evidently the $\sigma_{h}$ indicator is a more sensitive measure than $\sigma_{S}$.

\section{Results}

\begin{figure}[htb]
  \centering
  \includegraphics[width=8cm]{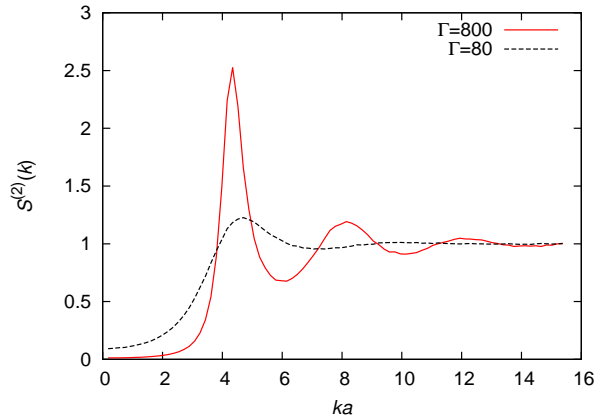}
  \caption{2-point static structure functions of the 3D Yukawa liquid at the $\Gamma$ values indicated, $\kappa=3$.}
  \label{ssf}
\end{figure}

We carried out MD-simulations for Yukawa-liquids with the system parameters of $\kappa=3$ and $\Gamma=\{80,800\}$. First, we present in Fig. \ref{ssf} the 2-point static structure functions for the parameters specified above. At these $\Gamma$ values the system is in the liquid phase, thus the system is isotropic, and the direction of the wave vector is irrelevant.

\begin{figure*}[floatfix,h!]
\begin{center}
\begin{tabular}{cc}
  \includegraphics[width=7.5cm]{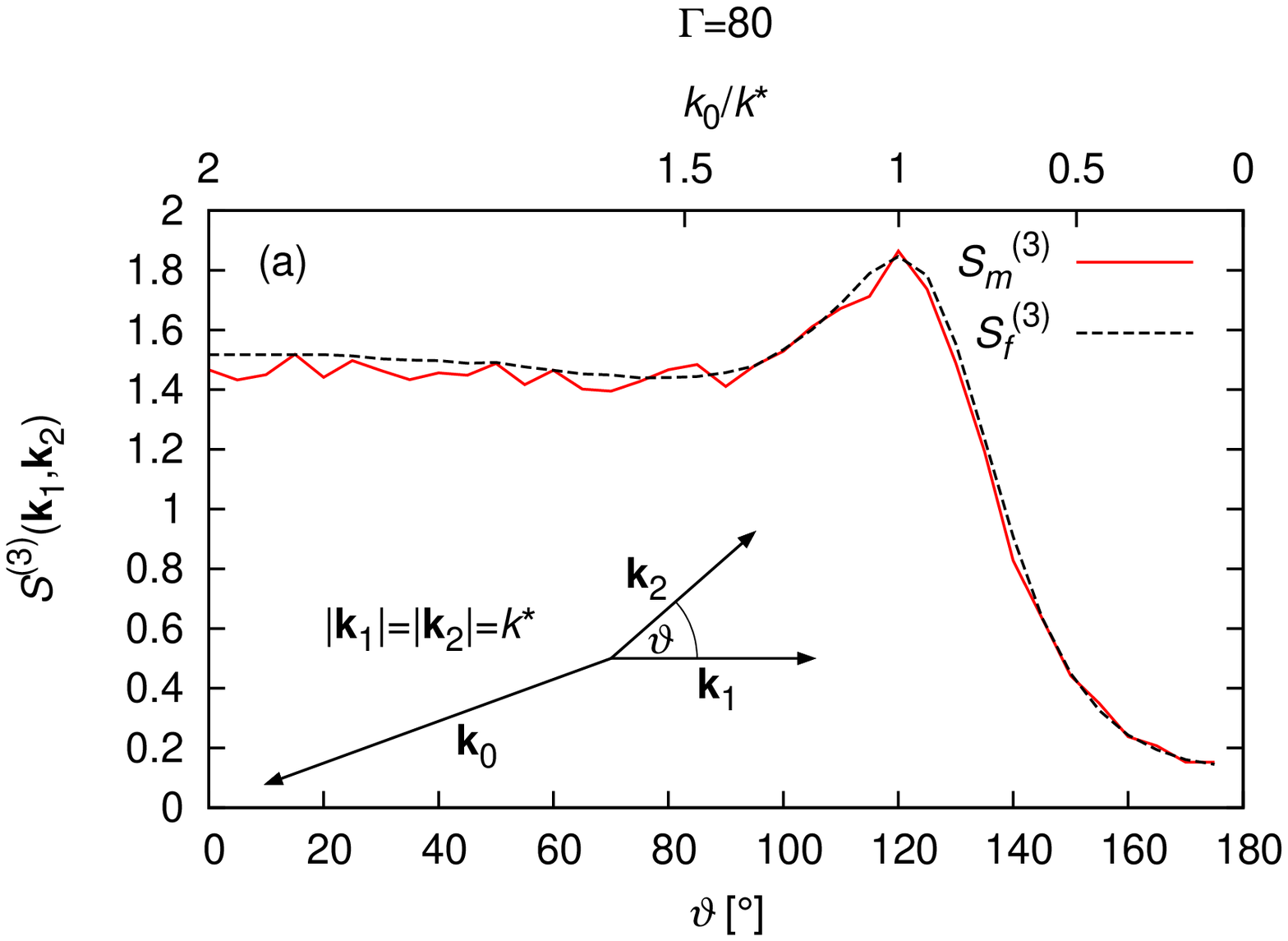}
 &
  \includegraphics[width=7.5cm]{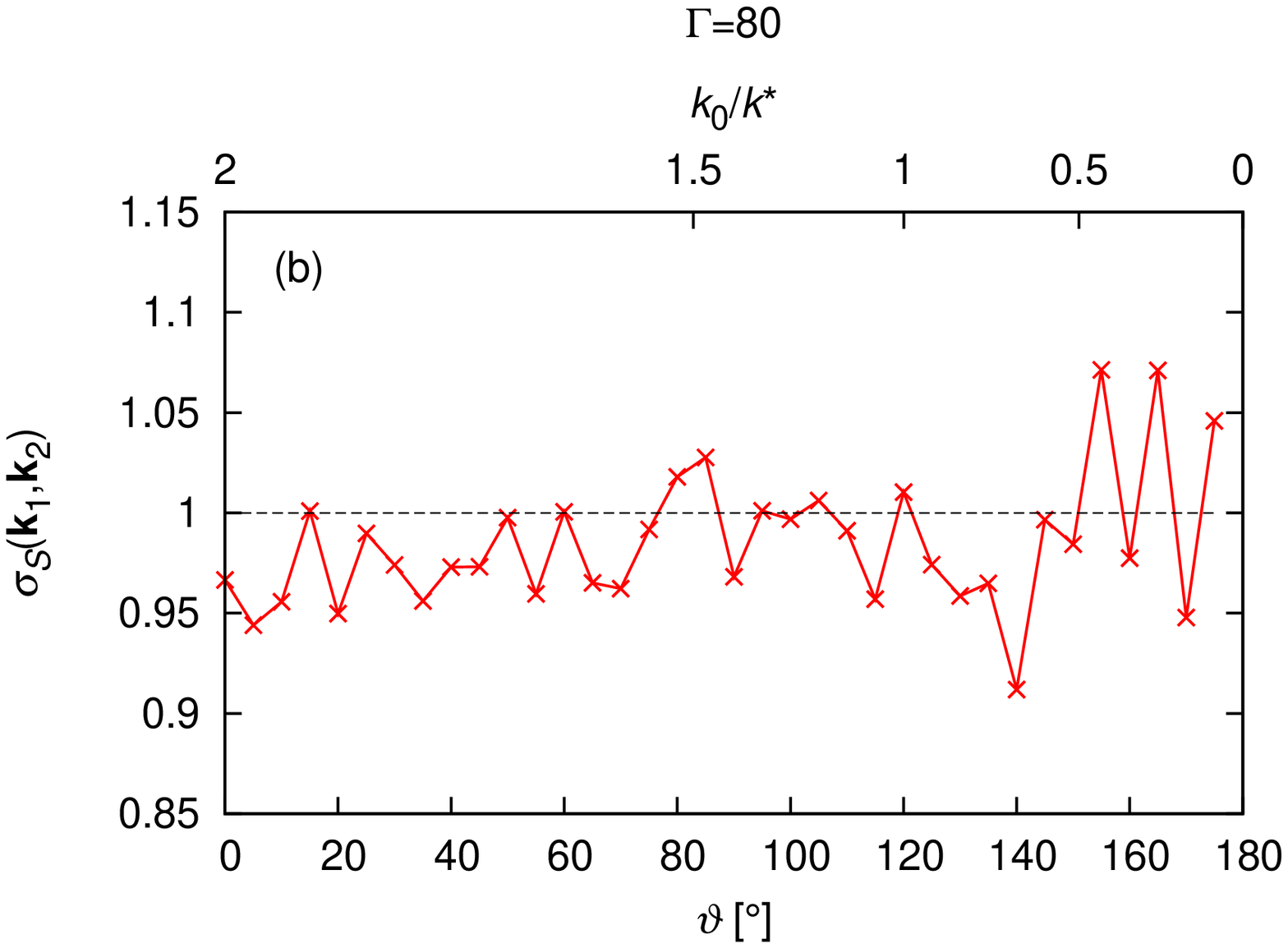}
 \\
\end{tabular}
\begin{tabular}{cc}
  \includegraphics[width=7.5cm]{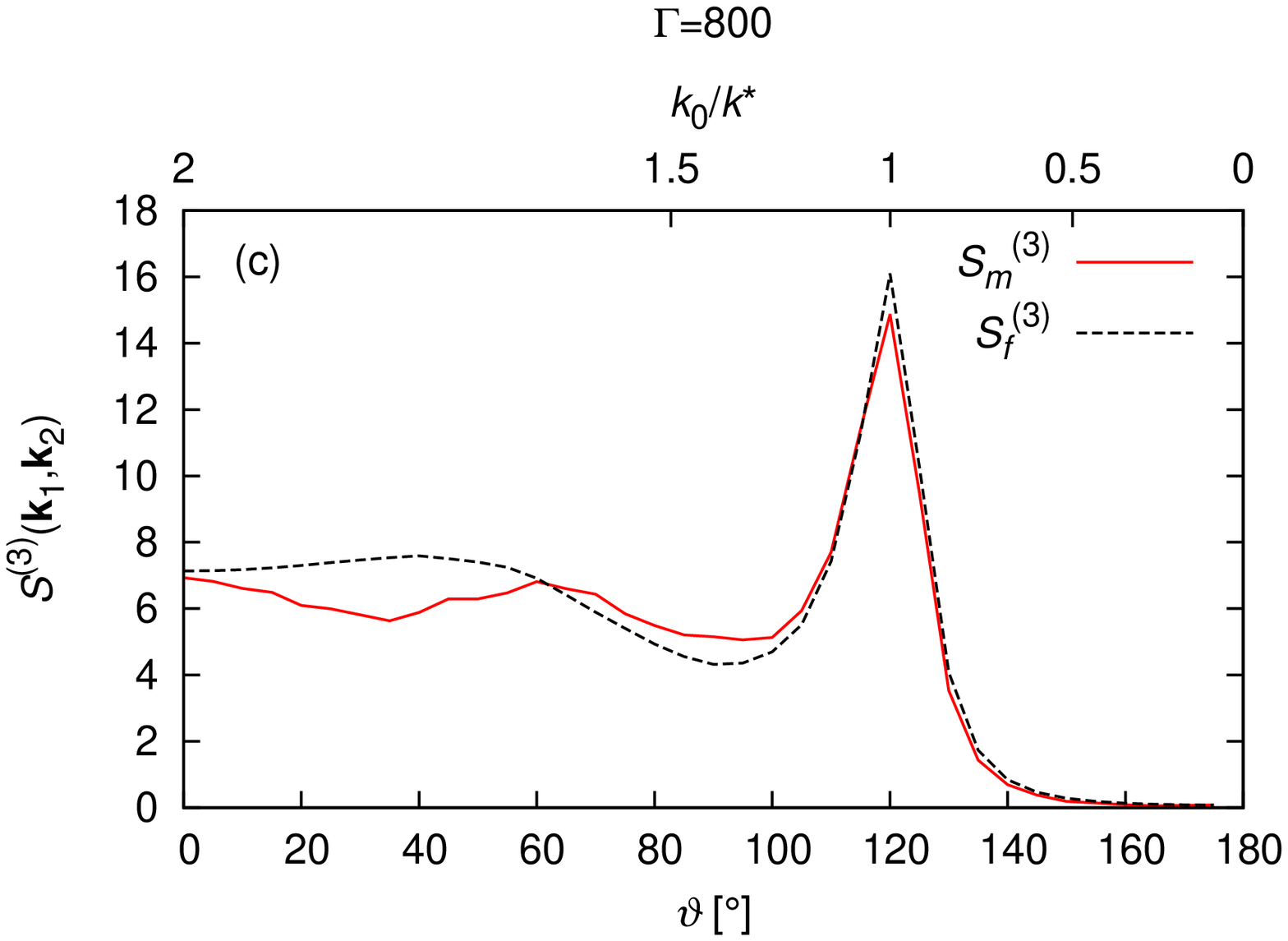}
 &
  \includegraphics[width=7.5cm]{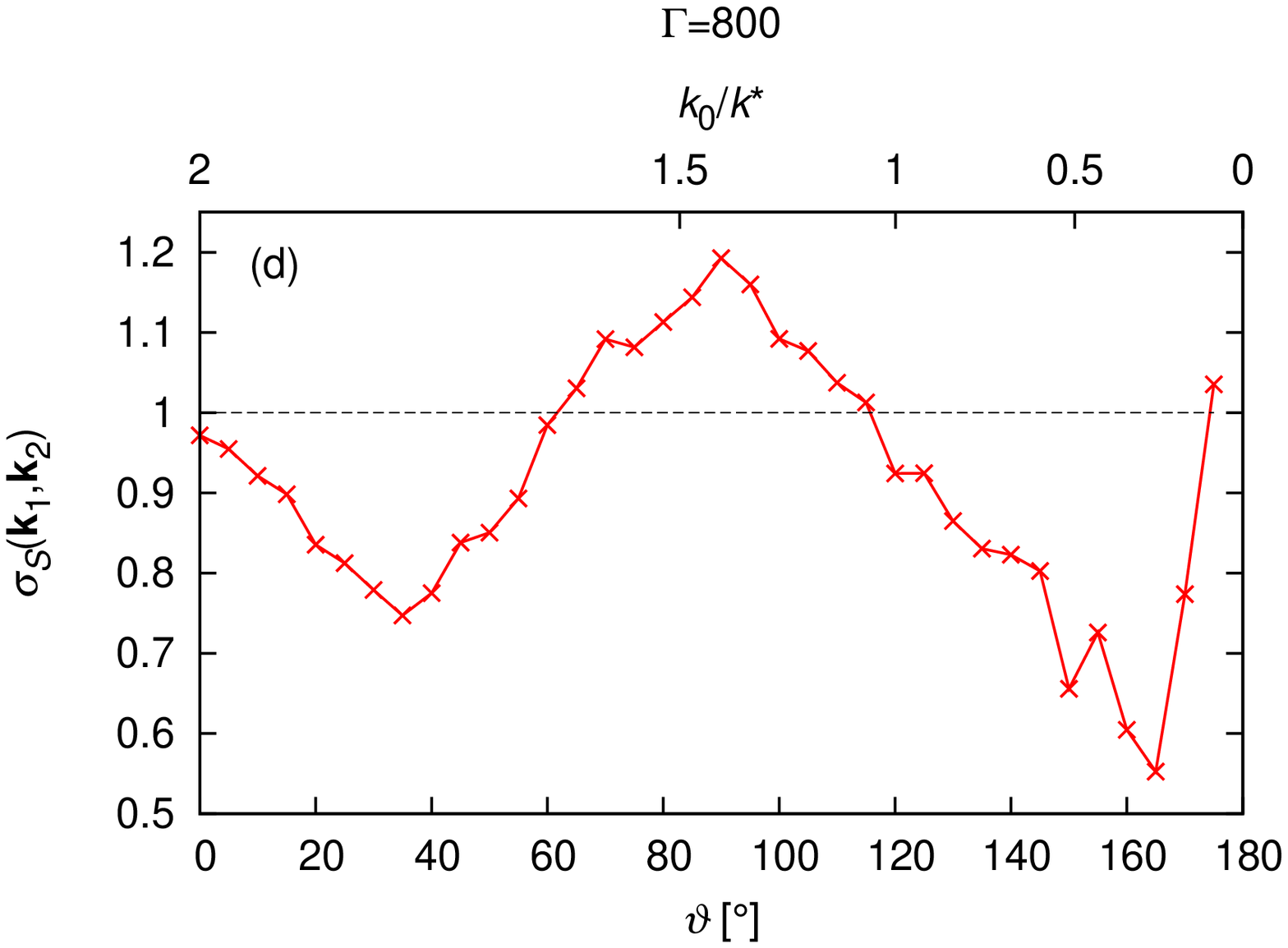}
 \\
\end{tabular}
\caption{(a,c) 3-point structure functions  $S^{(3)}(\textbf{k}_1,\textbf{k}_2)$ and (b,d) $\sigma_{S}(\textbf{k}_1,\textbf{k}_2)$ functions for isosceles triangles of the wave number vectors, for the $\Gamma$ values indicated and $\kappa=3$. $k^\ast a=4.7$ for $\Gamma=80$ and $k^\ast a=4.34$ for $\Gamma=800$, $k_{0}$ is as indicated on the upper scale. }
\label{S-angle}
\end{center}
\end{figure*}

\begin{figure*}[floatfix,h!]
\begin{center}
\begin{tabular}{cc}
  \includegraphics[width=7.5cm]{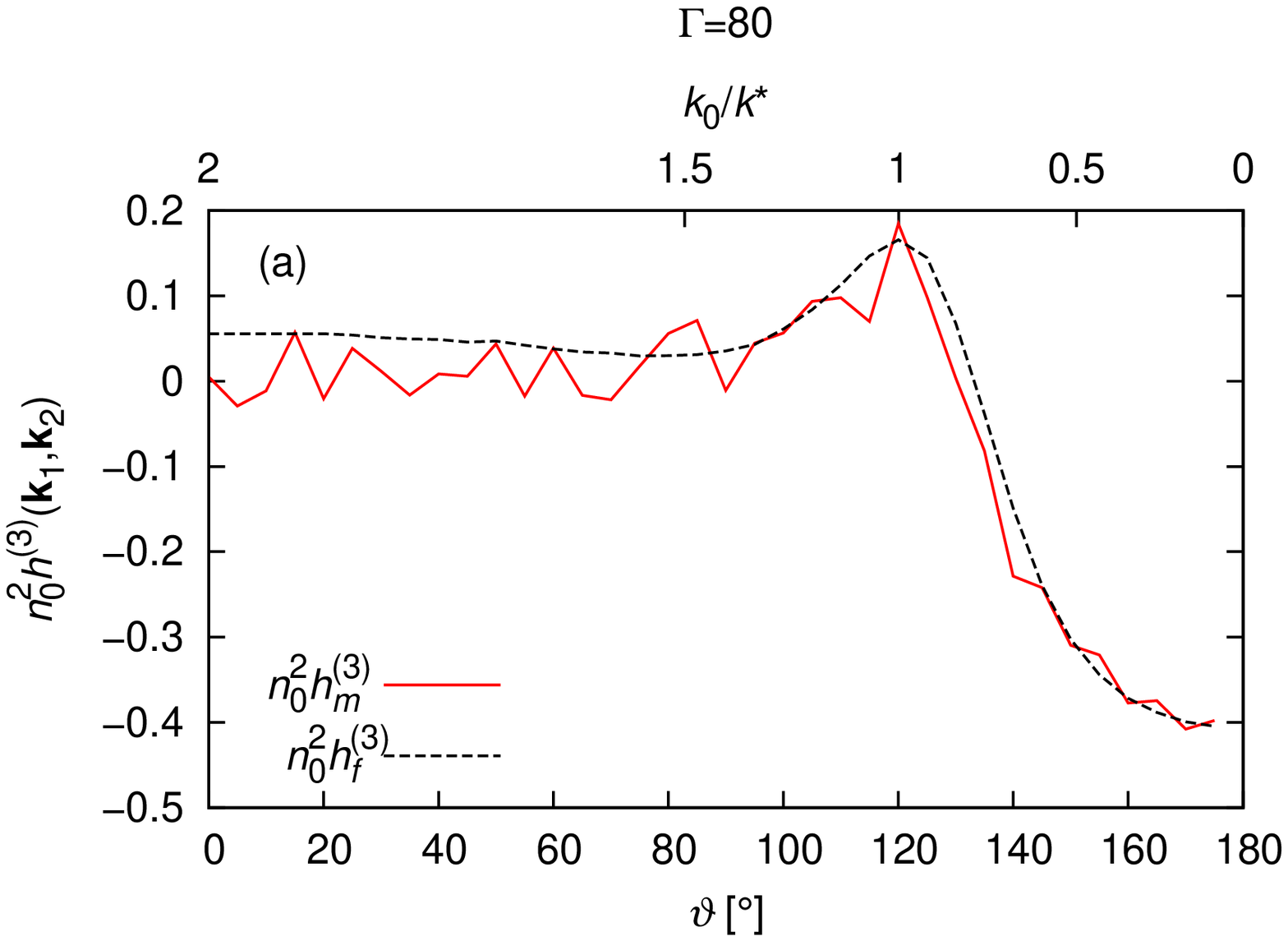}
 &
  \includegraphics[width=7.5cm]{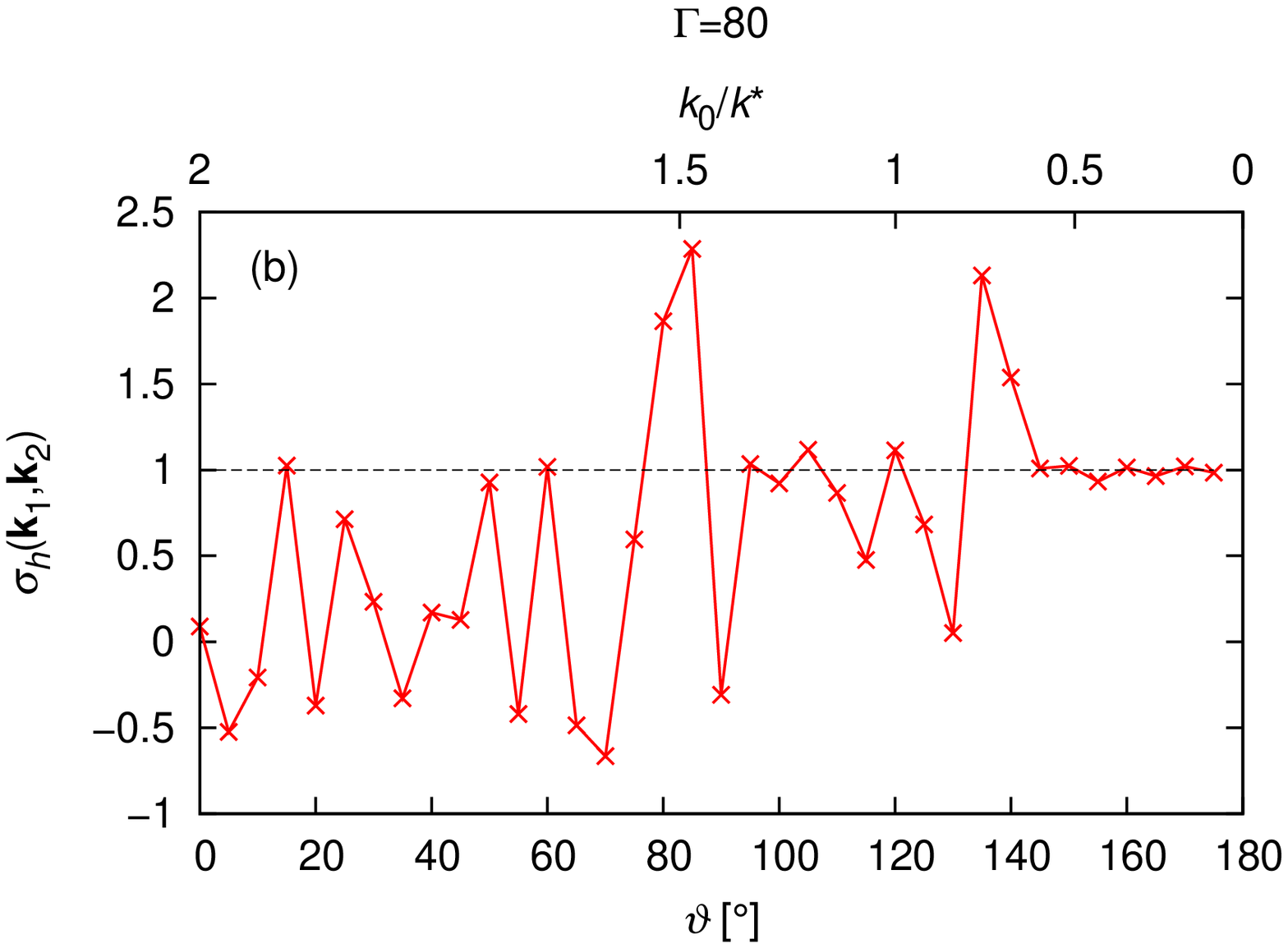}
 \\
\end{tabular}
\begin{tabular}{cc}
  \includegraphics[width=7.5cm]{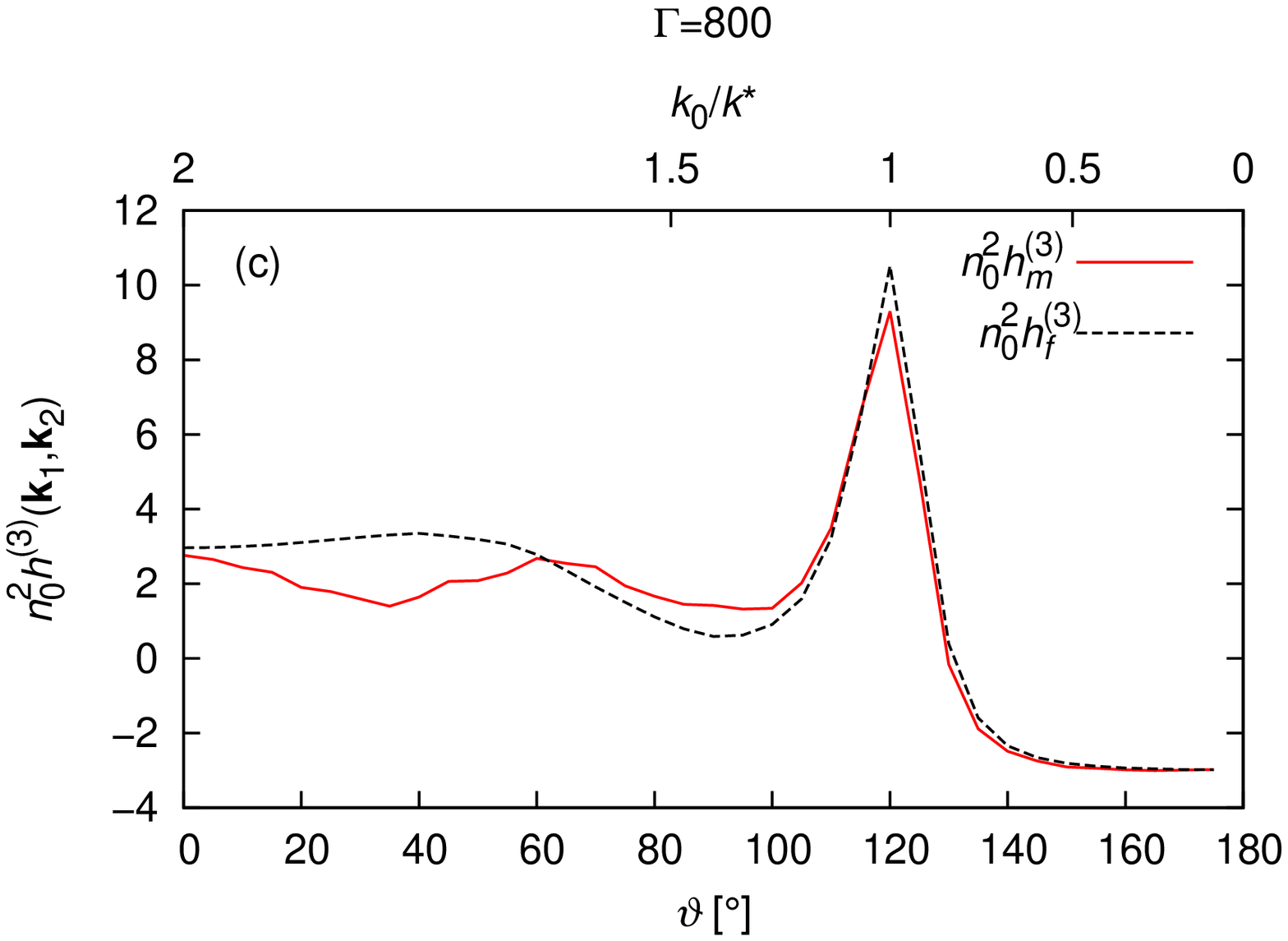}
 &
  \includegraphics[width=7.5cm]{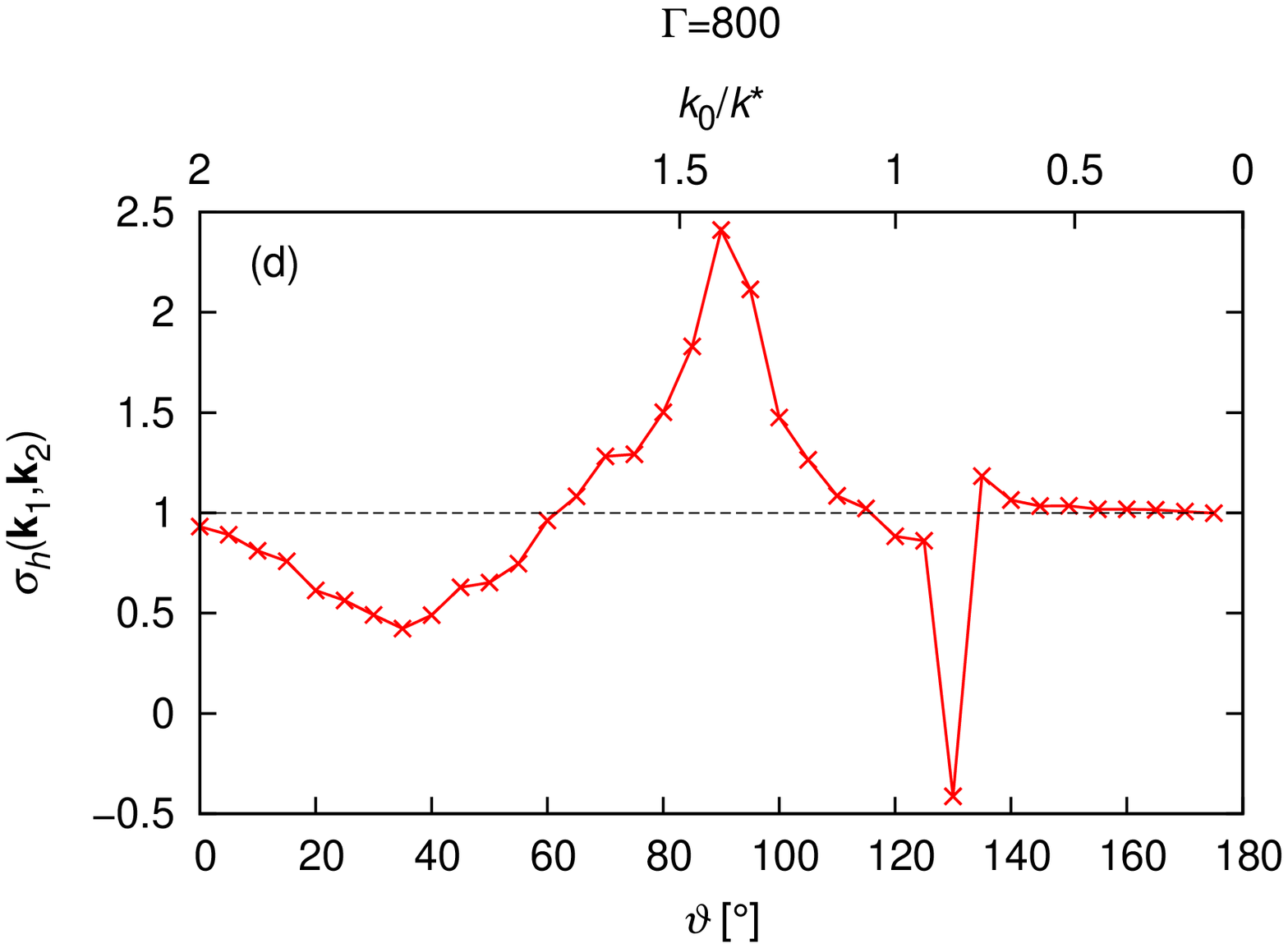}
 \\
\end{tabular}
\caption{(a,c) Triplet correlation functions $h^{(3)}(\textbf{k}_1,\textbf{k}_2)$ and (b,d) $\sigma_{h}(\textbf{k}_1,\textbf{k}_2)$ functions for isosceles triangles of the wave number vectors, for the $\Gamma$ values indicated and $\kappa=3$. $k^\ast a=4.7$ for $\Gamma=80$ and $k^\ast a=4.34$ for $\Gamma=800$, $k_{0}$ is as indicated on the upper scale. }
\label{h-angle}
\end{center}
\end{figure*}

First, we examine the case of the \textbf{isosceles triangles} with $|\textbf{k}_{1}|=|\textbf{k}_{2}|\equiv k^\ast$ and various angles $\vartheta$ between the vectors $\textbf{k}_{1}$ and $\textbf{k}_{2}$, where $k^\ast$ is the position of the main peak in the structure function $S^{(2)}(k)$ ($k^\ast a=26 k_{min}a = 4.7$ at $\Gamma=80$ and $k^\ast a=24 k_{min}a = 4.34$ at $\Gamma=800$). The angle $\vartheta$ changes between $0^\circ$ and $175^\circ$ with steps of 5$^\circ$. The results obtained for the 3-point structure functions and the triplet correlation functions are shown in Fig. \ref{S-angle} and Fig. \ref{h-angle}, respectively.

Figure \ref{S-angle}(a) displays the 3-point structure function for $\Gamma=80$. The ``measured'' data, $S^{(3)}_{m}(\textbf{k}_1,\textbf{k}_2)$ , (obtained directly from the simulation) and the data obtained from the 2-point structure function via relation (\ref{factorization}), $S^{(3)}_{f}(\textbf{k}_1,\textbf{k}_2)$, agree with each other within the noise level. (We note that the noise level in the determination of the 3-point structure function is significantly higher, compared to that in the computation of the 2-point structure function.) The ``quality of factorization'', as reflected by the data obtained for $\sigma_{S}(\textbf{k}_1,\textbf{k}_2)$ (see Fig. \ref{S-angle}(b)), is very good, the factorization of $S^{(3)}(\textbf{k}_1,\textbf{k}_2)$ in terms of 2-point $S(\textbf{k}_{1})$-s holds within a few percent.

At the higher coupling ($\Gamma=800$), however, Fig. \ref{S-angle}(c) indicates significant, even qualitative, differences between
$S^{(3)}_{m}(\textbf{k}_1,\textbf{k}_2)$ and $S^{(3)}_{f}(\textbf{k}_1,\textbf{k}_2)$, especially at angles $\vartheta < 100^\circ$. These deviations hinder the possibility of an approximate factorization, as indicated by the $\sigma_{S}(\textbf{k}_1,\textbf{k}_2)$ values shown in Fig. \ref{S-angle}(d). An approximate factorization can here both underestimate and overestimate $S^{(3)}(\textbf{k}_1,\textbf{k}_2)$, depending on $\vartheta$, by up to $\approx$ 40\%.

A similar study on the triplet correlation function, $h^{(3)}(\textbf{k}_1,\textbf{k}_2)$, is presented in Fig. \ref{h-angle}. Here, we find large differences between the measured $h^{(3)}_{m}(\textbf{k}_1,\textbf{k}_2)$ and the approximate $h^{(3)}_{f}(\textbf{k}_1,\textbf{k}_2)$ even at the lower coupling value of $\Gamma=80$. These deviations make the determination of $h^{(3)}(\textbf{k}_1,\textbf{k}_2)$ uncertain, as reflected by the $\sigma_{h}(\textbf{k}_1,\textbf{k}_2)$ data shown in Fig. \ref{h-angle}(b). At the higher coupling our data are less noisy and the $\sigma_{h}(\textbf{k}_1,\textbf{k}_2)$ is well-defined (see Fig. \ref{h-angle}(d)). (Note that a data point in $\sigma_{h}(\textbf{k}_1,\textbf{k}_2)$ at $\vartheta =135^\circ$ is ill-defined due to the zero crossing of $h^{(3)}_{f}(\textbf{k}_1,\textbf{k}_2)$.)

\begin{figure*}[floatfix,h!]
\begin{center}
\begin{tabular}{cc}
  \includegraphics[width=7.5cm]{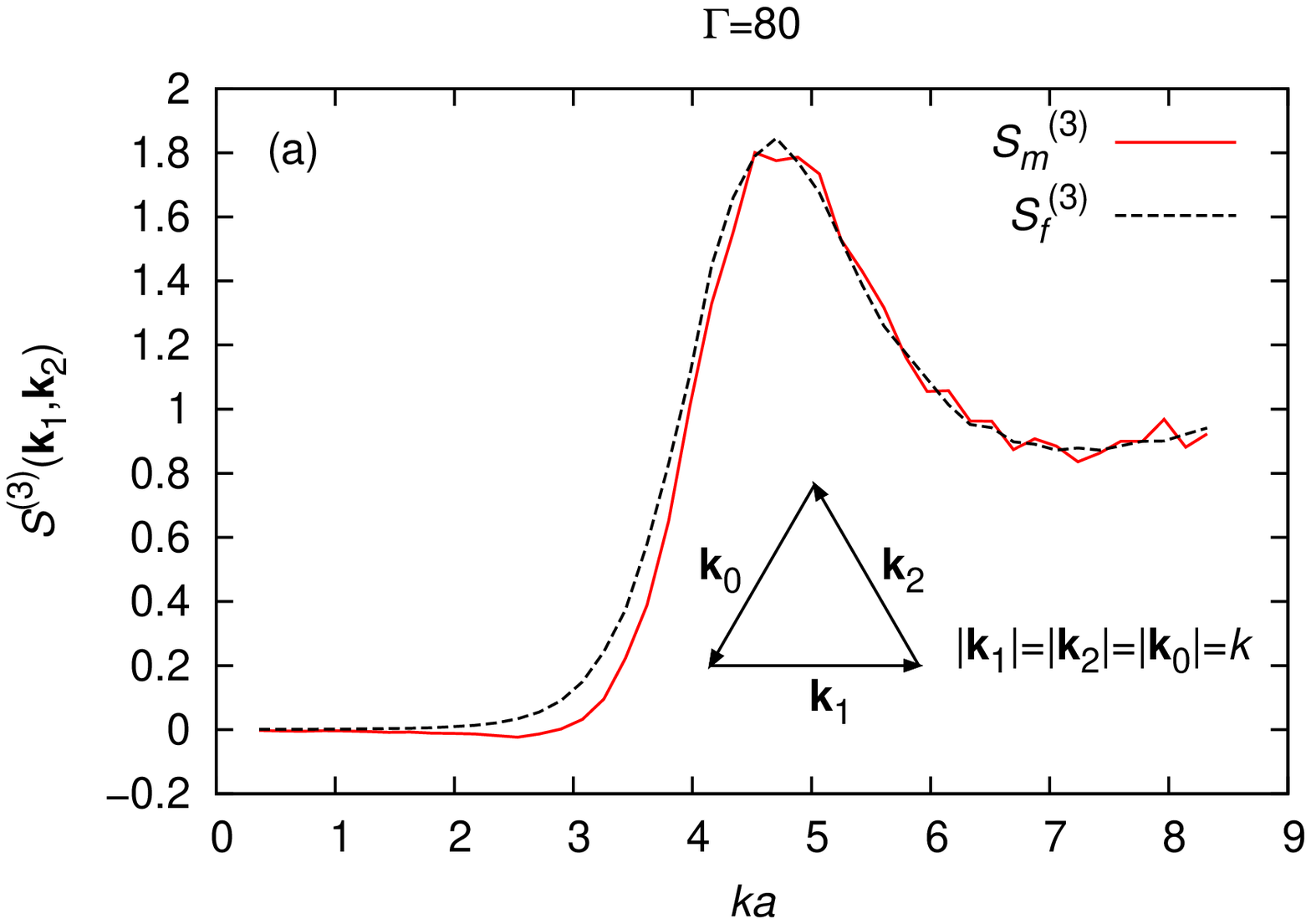}
 &
  \includegraphics[width=7.5cm]{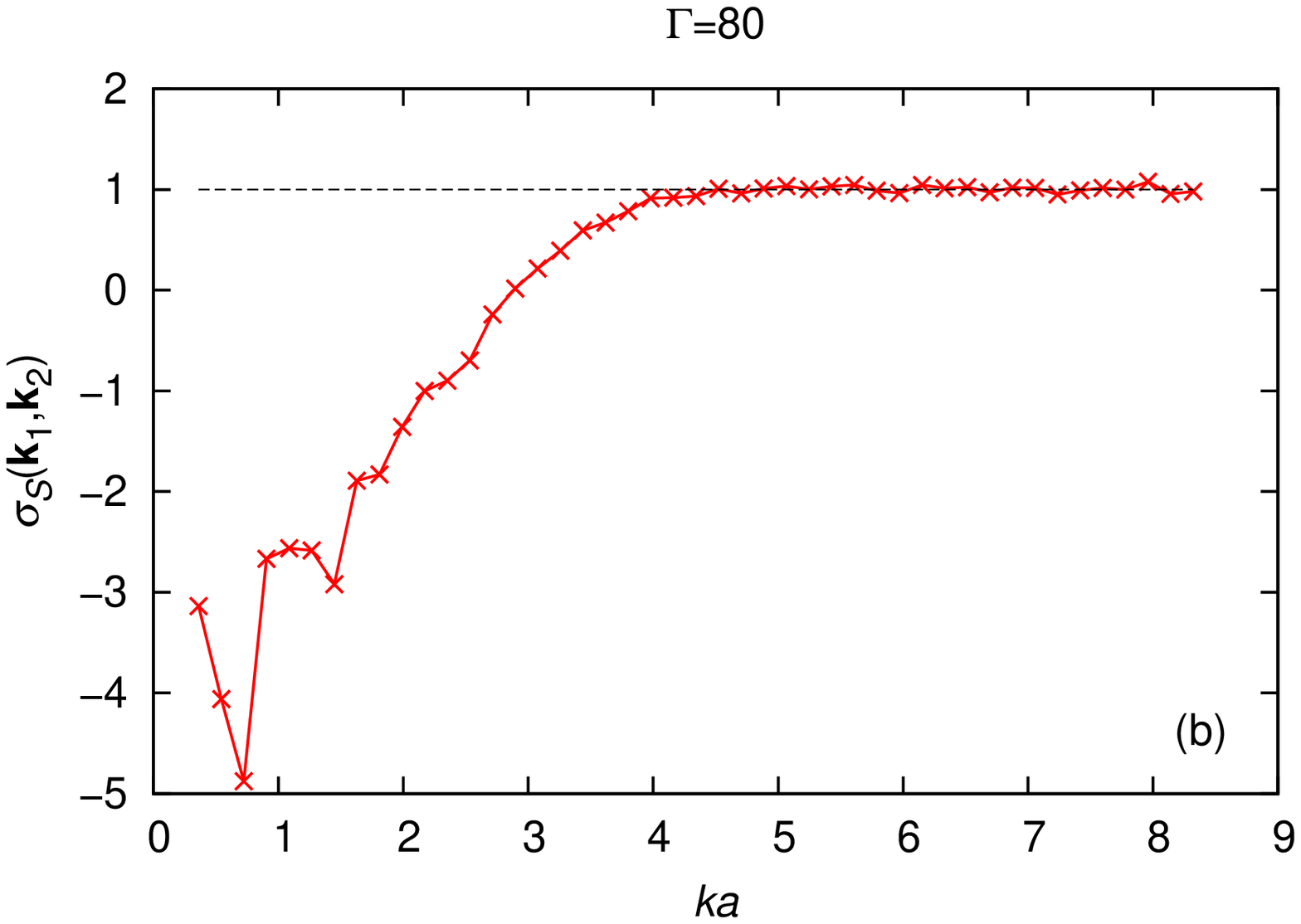}
 \\
\end{tabular}
\begin{tabular}{cc}
  \includegraphics[width=7.5cm]{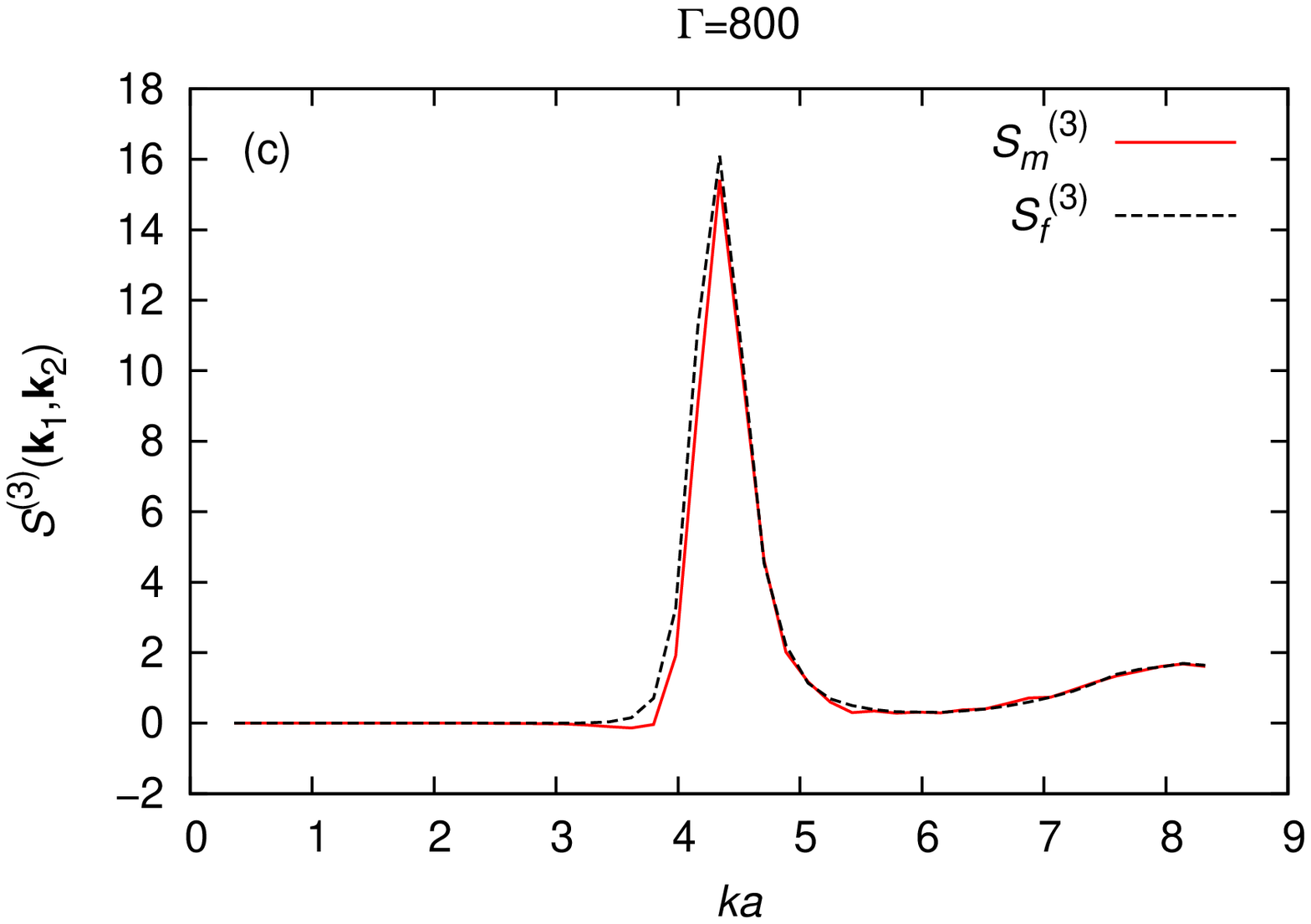}
 &
  \includegraphics[width=7.5cm]{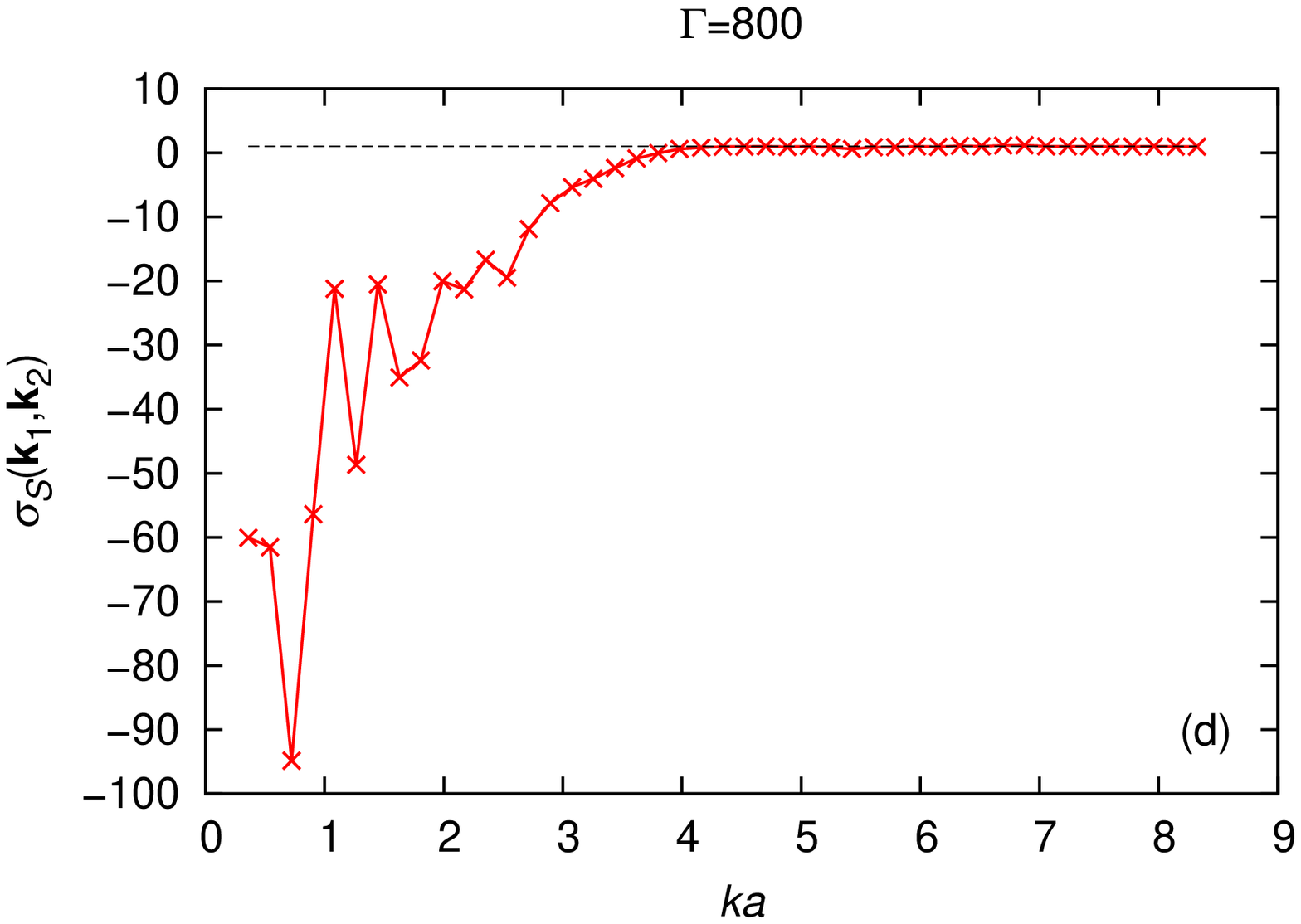}
 \\
\end{tabular}
\caption{(a,c) 3-point structure functions  $S^{(3)}(\textbf{k}_1,\textbf{k}_2)$ and (b,d) $\sigma_{S}(\textbf{k}_1,\textbf{k}_2)$ functions for equilateral triangles of the wave number vectors, $k\equiv |\textbf{k}_{1}|=|\textbf{k}_{2}|=|\textbf{k}_{0}|$, for the $\Gamma$ values indicated. The negative values of $\sigma_{S}$ are caused by the negative values of $S^{(3)}_{m}$, as $S^{(3)}_{f}$ is positive in the whole interval of $ka$ examined. $\kappa=3$.}
\label{S-equilateral-k3}
\end{center}
\end{figure*}

\begin{figure*}[floatfix,h!]
\begin{center}
\begin{tabular}{cc}
  \includegraphics[width=7.5cm]{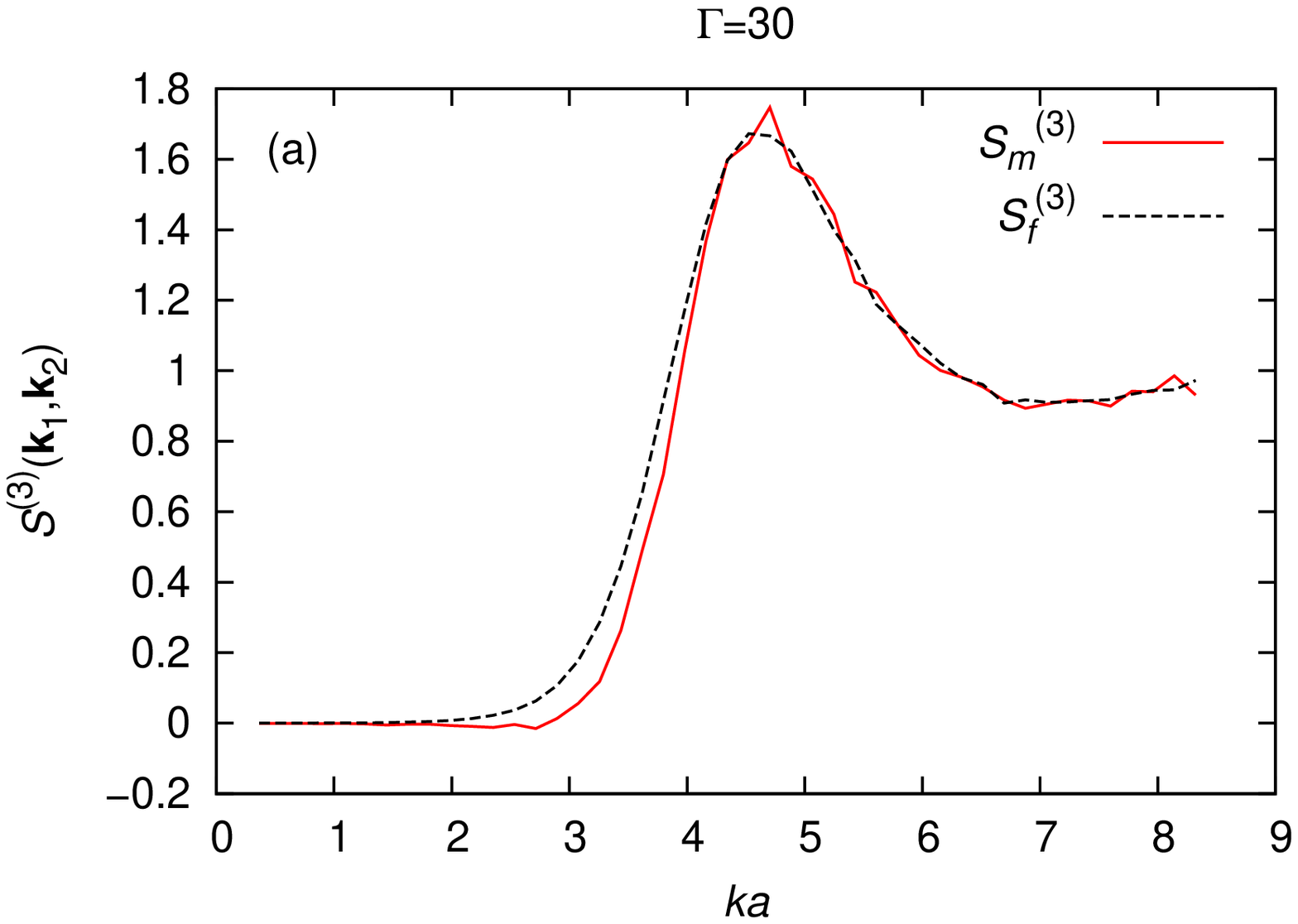}
 &
  \includegraphics[width=7.5cm]{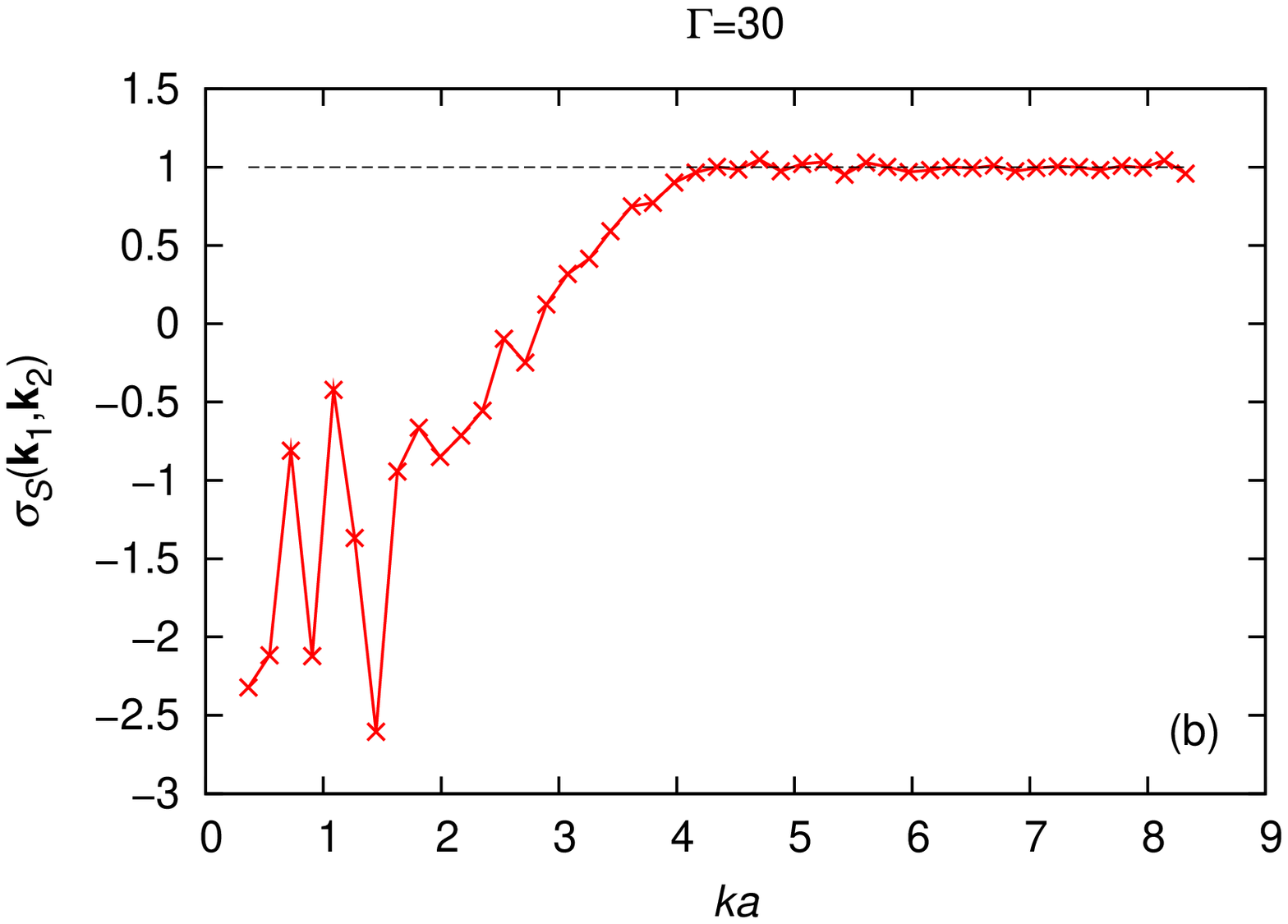}
 \\
\end{tabular}
\begin{tabular}{cc}
  \includegraphics[width=7.5cm]{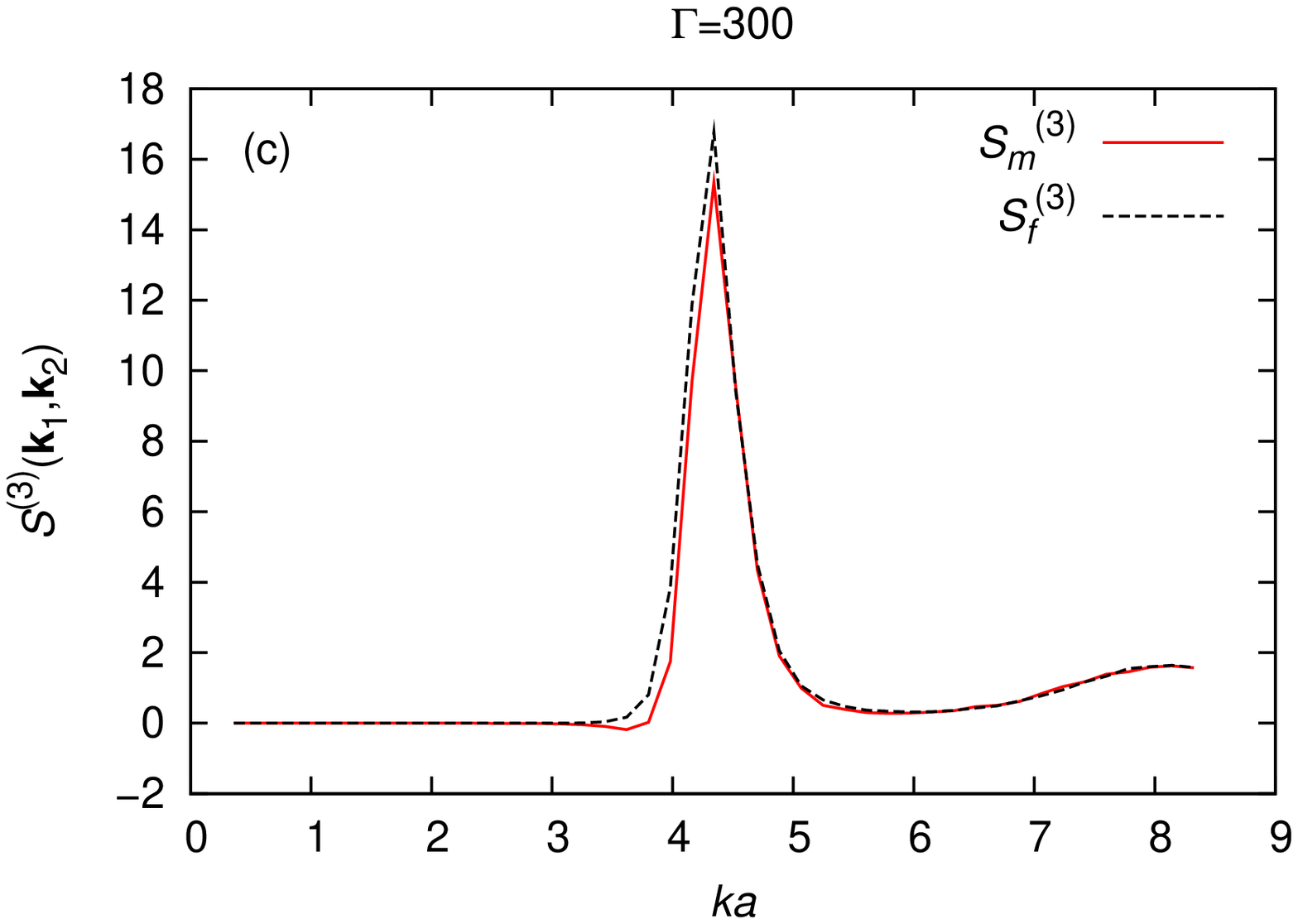}
 &
  \includegraphics[width=7.5cm]{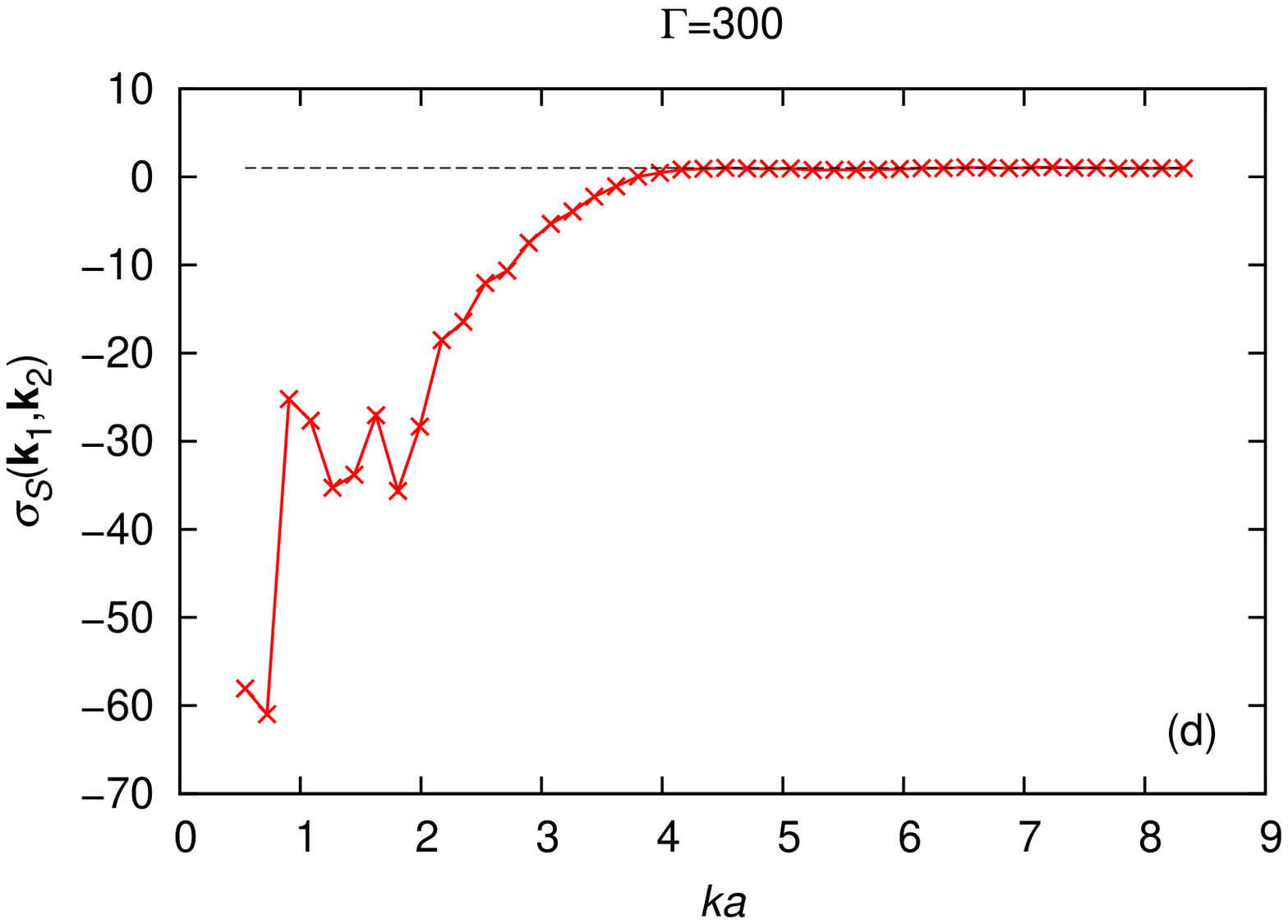}
 \\
\end{tabular}
\caption{(a,c) 3-point structure functions  $S^{(3)}(\textbf{k}_1,\textbf{k}_2)$ and (b,d) $\sigma_{S}(\textbf{k}_1,\textbf{k}_2)$ functions for equilateral triangles of the wave number vectors, $k\equiv |\textbf{k}_{1}|=|\textbf{k}_{2}|=|\textbf{k}_{0}|$, for the $\Gamma$ values indicated and $\kappa=2$. The negative values of $\sigma_{S}$ are caused by the negative values of $S^{(3)}_{m}$, as $S^{(3)}_{f}$ is positive in the whole interval of $ka$ examined.}
\label{S-equilateral-k2}
\end{center}
\end{figure*}

\begin{figure*}[floatfix,h!]
\begin{center}
\begin{tabular}{cc}
  \includegraphics[width=7.5cm]{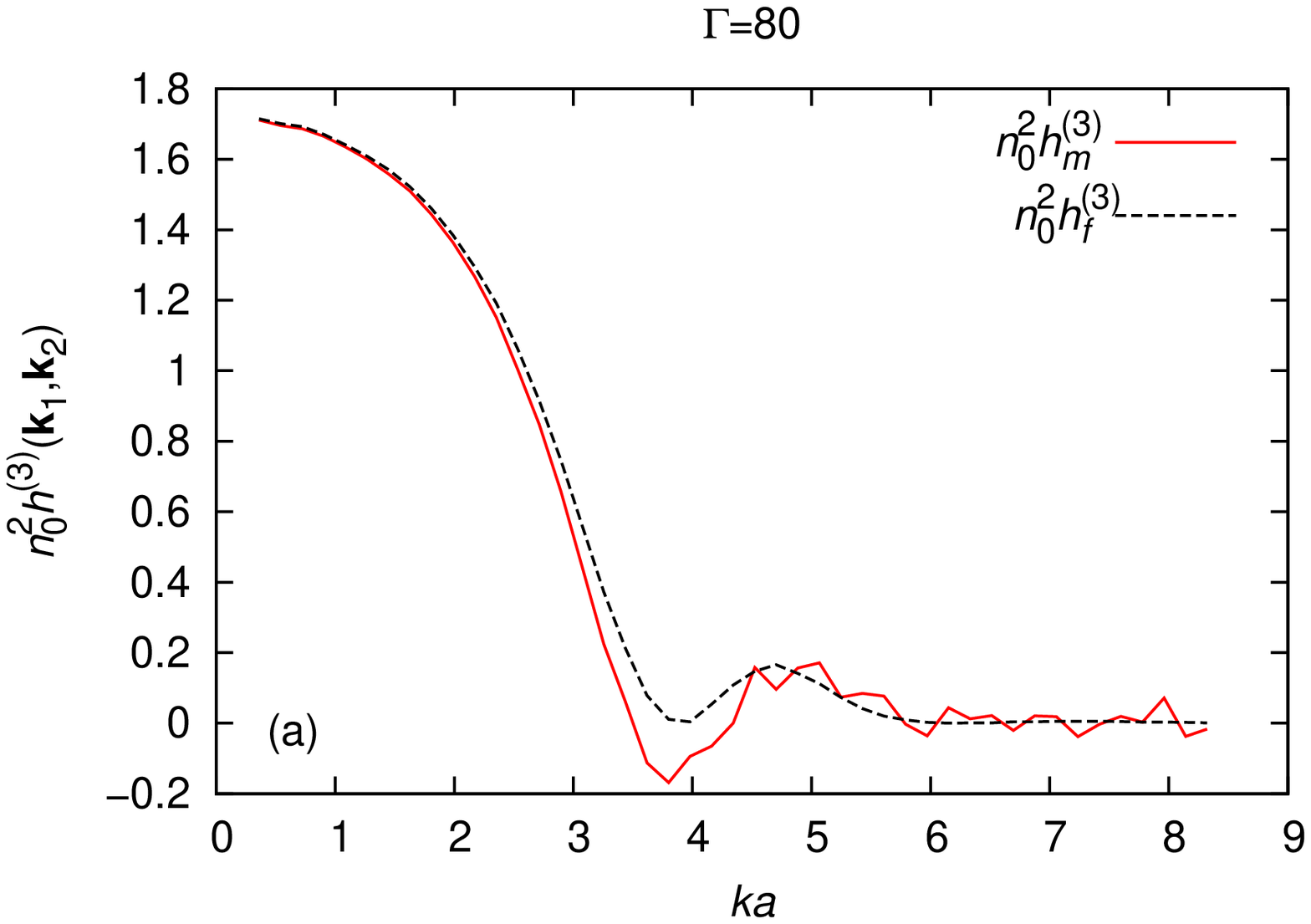}
 &
  \includegraphics[width=7.5cm]{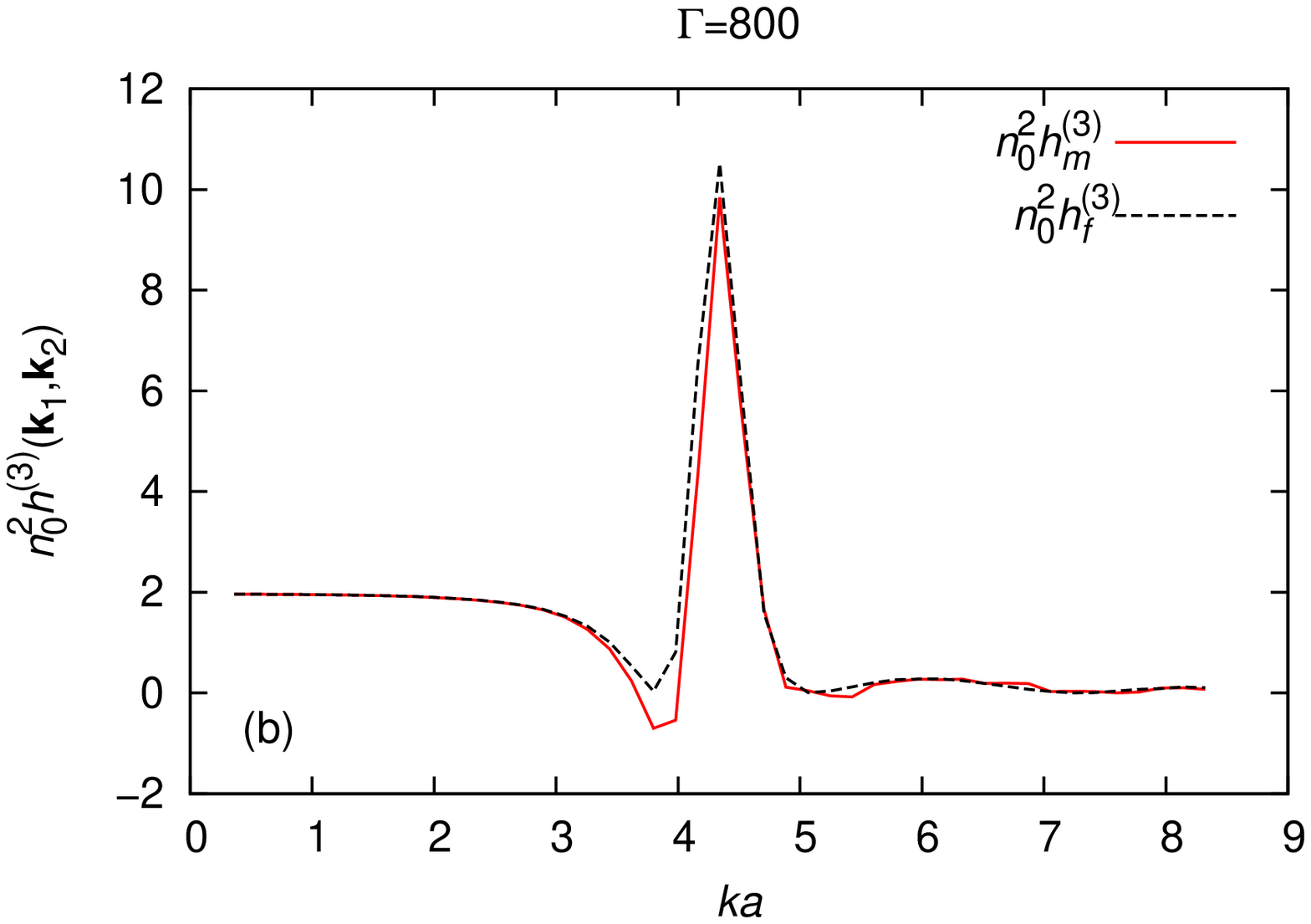}
 \\
\end{tabular}
\caption{Triplet correlation functions $h^{(3)}(\textbf{k}_1,\textbf{k}_2)$ for equilateral triangles of wave number vectors, $k\equiv |\textbf{k}_{1}|=|\textbf{k}_{2}|=|\textbf{k}_{0}|$, for the $\Gamma$ values indicated and $\kappa=3$.}
\label{h-equilateral-k3}
\end{center}
\end{figure*}

\begin{figure*}[floatfix,h!]
\begin{center}
\begin{tabular}{cc}
  \includegraphics[width=7.5cm]{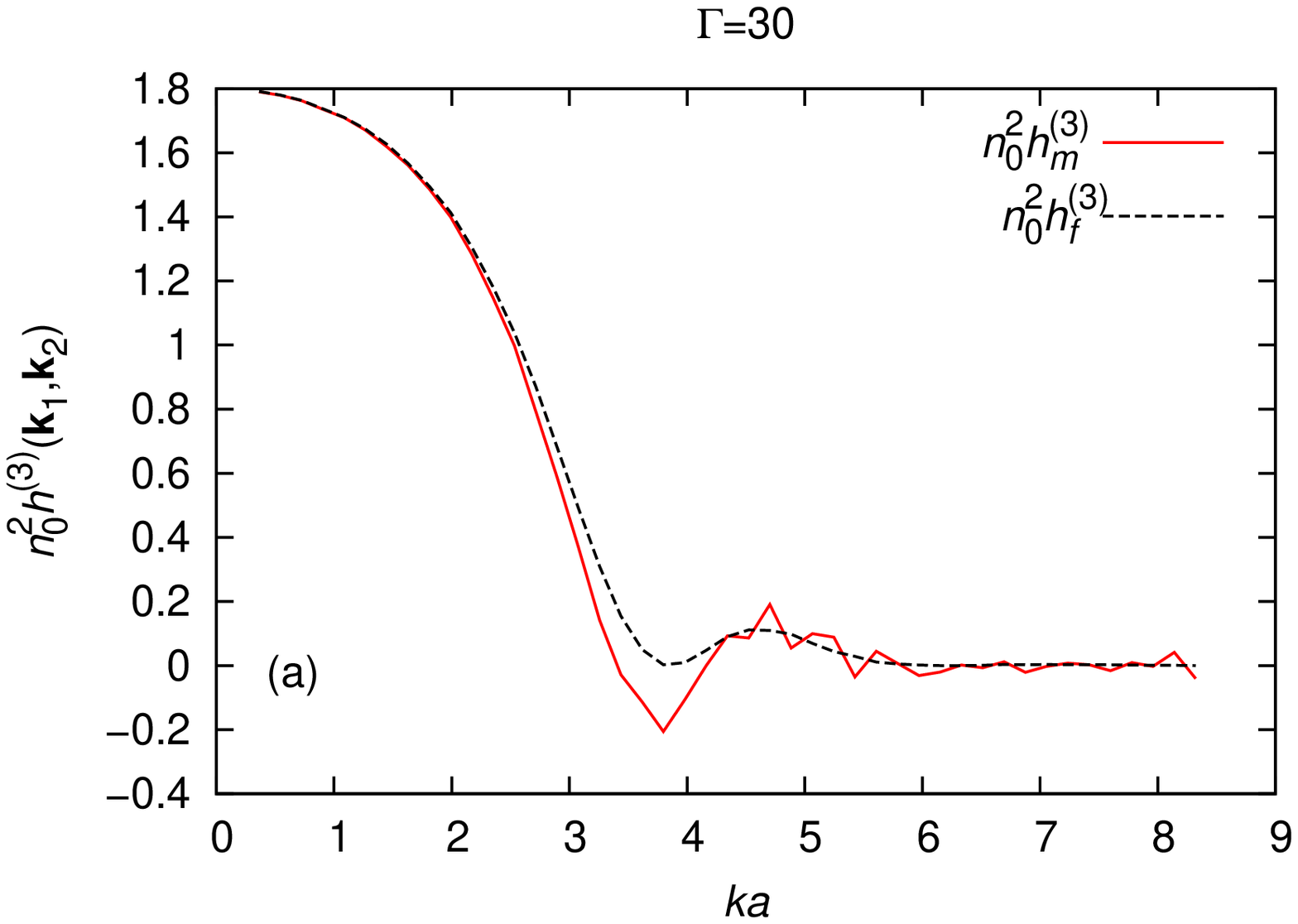}
 &
  \includegraphics[width=7.5cm]{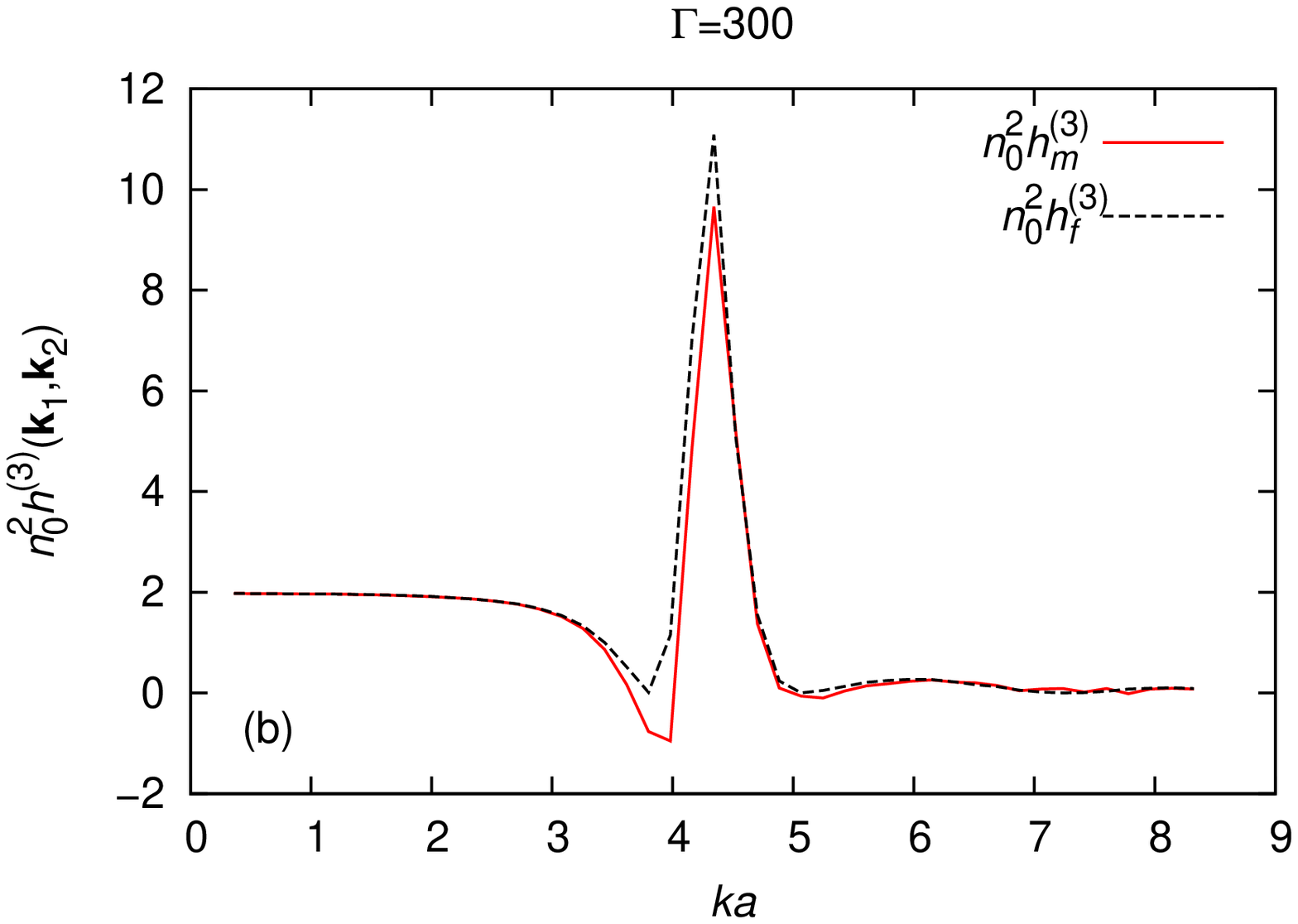}
 \\
\end{tabular}
\caption{Triplet correlation functions $h^{(3)}(\textbf{k}_1,\textbf{k}_2)$ for equilateral triangles of wave number vectors, $k\equiv |\textbf{k}_{1}|=|\textbf{k}_{2}|=|\textbf{k}_{0}|$, for the $\Gamma$ values indicated and $\kappa=2$.}
\label{h-equilateral-k2}
\end{center}
\end{figure*}

Next we consider the case of the \textbf{equilateral triangles} ($\vartheta=120^\circ$) with variable side lengths $k$. For this setting we have carried out computations for $\kappa=3$ and 2. The coupling values for the systems characterized by $\kappa=2$ ($\Gamma$ = 300 and 30) have been chosen to result in the same $\Gamma / \Gamma_{\rm m}$, as in the case of $\kappa=3$. Here, $\Gamma_{\rm m}$ is the coupling that corresponds to melting, its values are taken from \cite{Hamaguchi}.
(Unfortunately we are not able to scan a wide range of screening values, due to the excessive computational requirements, especially at lower values of $\kappa$.) The value of $k$ changes in both cases between $2k_{min}$ and $46k_{min}$ with a step of $k_{min}$. The results obtained for the 3-point structure functions are shown in Fig. \ref{S-equilateral-k3} and Fig. \ref{S-equilateral-k2} for the different $\kappa$ values, while the corresponding triplet correlation functions are displayed in Fig. \ref{h-equilateral-k3} and Fig. \ref{h-equilateral-k2}.

In the case of the 3-point structure functions (Figs. \ref{S-equilateral-k3} and \ref{S-equilateral-k2}) we observe similar shapes and a good quality of the approximate factorization at high wave numbers, $k a > 4$. At smaller wave numbers, however, although the shapes of $S^{(3)}_{m}(\textbf{k}_1,\textbf{k}_2)$ and $S^{(3)}_{f}(\textbf{k}_1,\textbf{k}_2)$ are very similar, we observe a strong breakdown of the approximate factorization, as indicated by the large negative values of $\sigma_{S}(\textbf{k}_1,\textbf{k}_2)$ (see panels (b) and (d) in Figs. \ref{S-equilateral-k3}  and Figs. \ref{S-equilateral-k2}). These values originate from the negative values of $S^{(3)}_{m}(\textbf{k}_1,\textbf{k}_2)$ below $k a \approx 4$. Obviously, negative values cannot be generated by the approximate factorization, as  $S(\textbf{k}_{1}) > 0$ for all wave numbers. Similarly to the 3-point structure function for the given setting of the wave number vectors, the triplet correlation function $h^{(3)}(\textbf{k}_1,\textbf{k}_2)$ also acquires negative values below $k a \approx 4$ (see Figs.~\ref{h-equilateral-k3} and \ref{h-equilateral-k2}), and the approximate factorization clearly cannot work in this domain.  The occurrence of a negative domain in $S^{(3)}$ is known from studies on different systems, however the {\it physical background} of this behavior is less clear. A changing sign of $S^{(3)}$ implies a changing sign of the second order static density response of the system to external potential energy perturbations, which can be directly observed in MD simulations, as we demonstrate below.

\begin{figure}[htb]
  \centering
  \includegraphics[width=8cm]{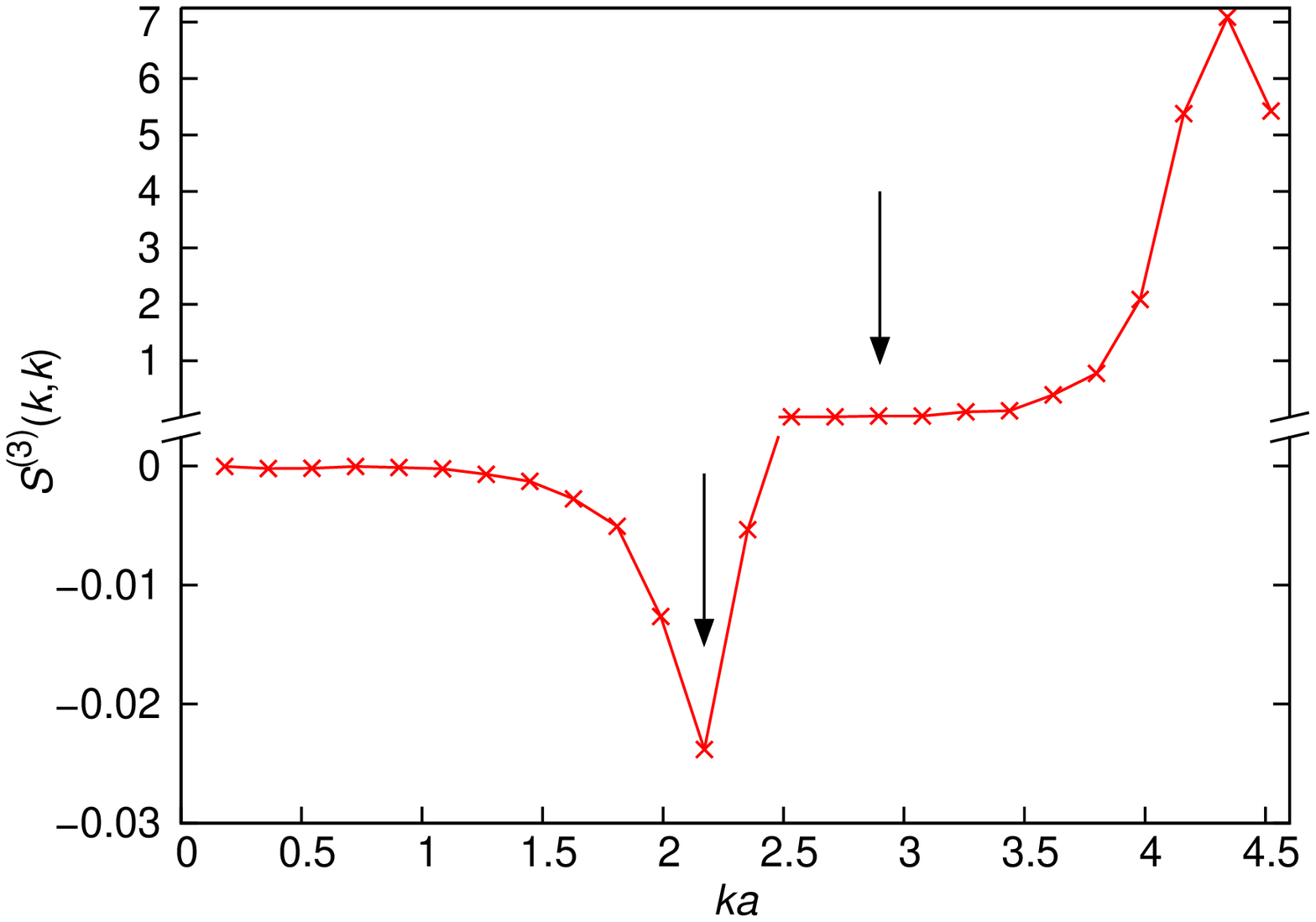}
  \caption{3-point static structure function for $\textbf{k}_{1} = \textbf{k}_{2} \equiv \textbf{k}$, for $\Gamma=800$ and $\kappa=3$. The two wave number values examined via the perturbation simulation are indicated by arrows.}
  \label{diagonalskk}
\end{figure}

We recall the connection between the static structure functions and the density response functions \cite{Magyar}. The static linear FDT connects the linear response function with the static 2-point structure function:
\begin{equation}
\hat{\chi}^{(1)}(\textbf{k}_1)=-\beta n_{0} S^{(2)}(\textbf{k}_1),
\label{eq:fdt1}
\end{equation}
while the static quadratic FDT establishes a relationship between the quadratic response function and the 3-point static structure function:
\begin{equation}
\hat{\chi}^{(2)}(\textbf{k}_{1},\textbf{k}_{2}) = \frac{\beta^{2}n_{0}}{2} S^{(3)}(\textbf{k}_{1},\textbf{k}_{2}),
\label{eq:fdt2}
\end{equation}
where $\beta = 1 / k_{\rm B}T$. According to the second relationship, the quadratic response function, and thus the quadratic part of the density response has to change sign when the 3-point static structure function changes sign. To confirm this expected behavior we determine $S^{(3)}(\textbf{k}_{1},\textbf{k}_{2})$ with $\textbf{k}_{1}  = \textbf{k}_{2}  \equiv \textbf{k}$, at $\Gamma = 800$ and $\kappa = 3$ via the simulation as described above and via a perturbation simulation approach that computes the density response of the system in the presence of an external potential energy, of the form \cite{Magyar}:
\begin{equation}
\hat{\Phi}(\textbf{r}) = C f_{0} \cos(\textbf{k}\cdot\textbf{r}).
\label{pot1}
\end{equation}
In this form $f_0$ is the amplitude of the perturbation, while $C$ is an additional parameter that ensures that the maximum of the external force at $f_0=1$ equals the force acting between two charged particles separated by a distance $r=a$. In this type of  simulations the external potential energy is applied from the initialization, measurement of the density profile starts after the equilibration of the system.

According to \cite{Magyar}, the first- and second-order density perturbations are given by:
\begin{eqnarray}
\langle\tilde{n}(\textbf{r})\rangle^{(1)}&=&C f_{0}\hat{\chi}^{(1)}(\textbf{k})\cos(\textbf{k}\cdot\textbf{r}), \label{1rend}\\
\langle\tilde{n}(\textbf{r})\rangle^{(2)}&=&\frac{C^{2}f_{0}^{2}}{2}\Big[\hat{\chi}^{(2)}(\textbf{k},-\textbf{k})+\hat{\chi}^{(2)}(\textbf{k},\textbf{k})\cos(2\textbf{k}\cdot\textbf{r})\Big]. \label{2rend}
\end{eqnarray}
The first-order density response follows the harmonic perturbation, via the link provided by the linear response function. The second-order density response contains (i) a ``DC contribution'' $\hat{\chi}^{(2)}(\textbf{k},-\textbf{k})$, for which ${\bf k}_0 = 0$, so it belongs to the domain of singular behavior \cite{Magyar}, and (ii) a term, $\hat{\chi}^{(2)}(\textbf{k},\textbf{k})\cos(2\textbf{k}\cdot\textbf{r})$  that is proportional to the quadratic response function.

\begin{figure*}[floatfix,h!]
\begin{center}
\begin{tabular}{cc}
  \includegraphics[width=7.5cm]{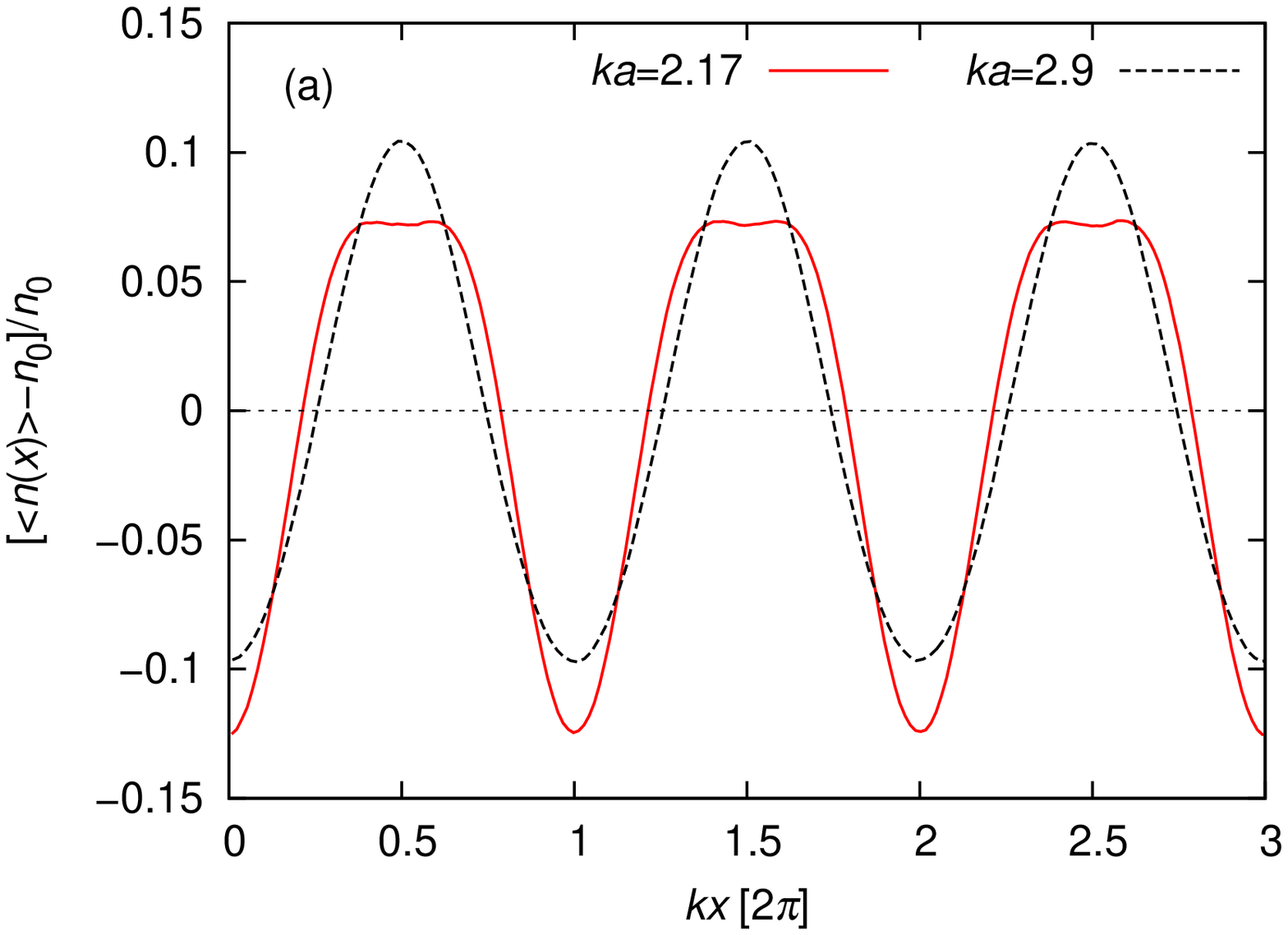}
 &
  \includegraphics[width=7.5cm]{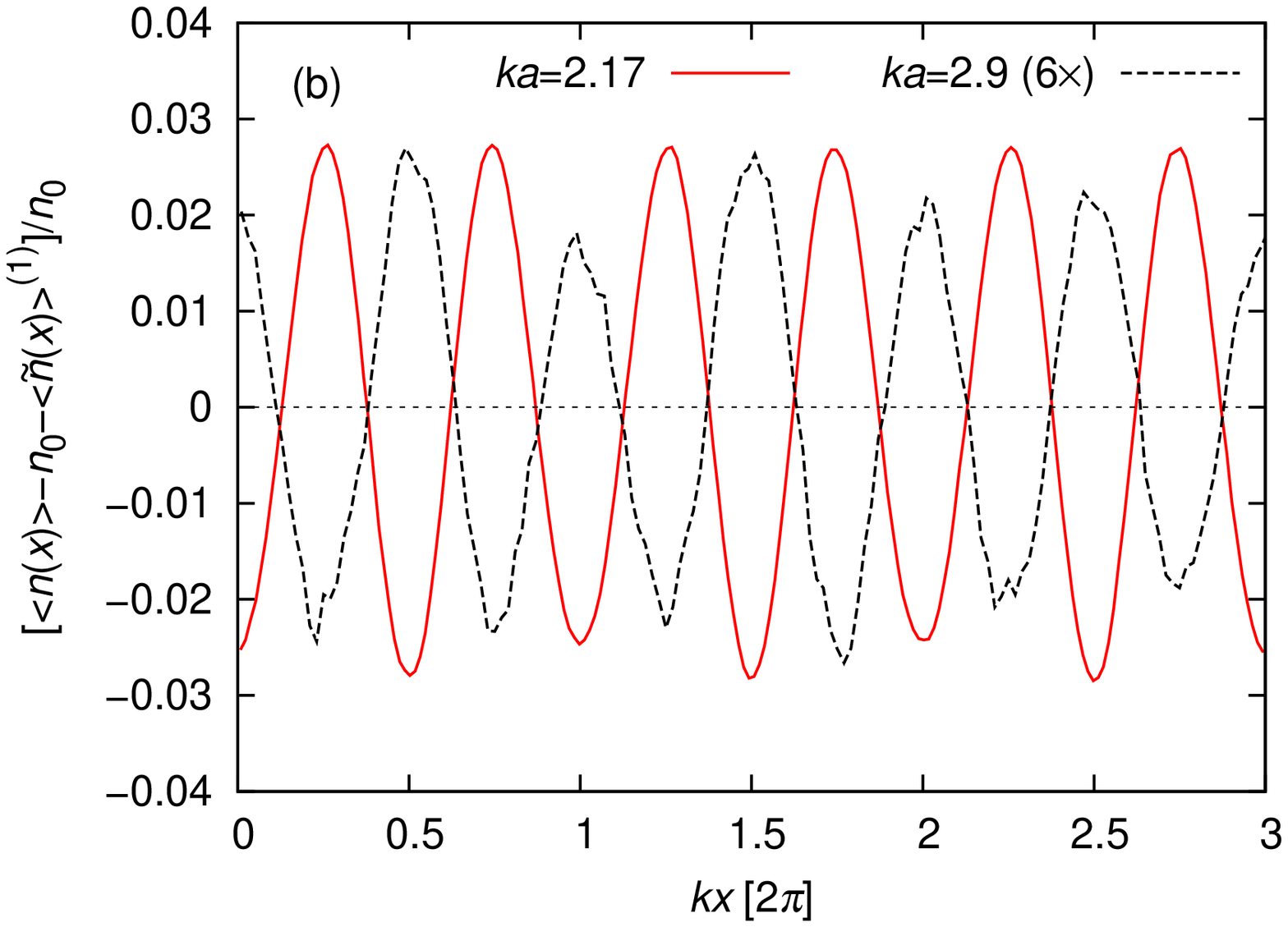}
 \\
\end{tabular}
\caption{(a) Total density response induced by a harmonic external potential energy perturbation with wave numbers $k a = 12 k_{\rm min} a=2.17$ and $k a = 16 k_{\rm min} a=2.9$. (The average density, $n_0$, is removed.) (b) Density profile resulting after removing the first-order response. We observe that the profiles oscillate in anti-phase, indeed confirming the sign change of the quadratic density response. $\Gamma=800$, $\kappa=3$.}
\label{response}
\end{center}
\end{figure*}

Based on the 3-point structure function $S^{(3)}(k,k)$, shown in figure \ref{diagonalskk} for $\Gamma = 800$ and $\kappa = 3$, we carry out the simulations with the perturbation method for $k a = 12 k_{\rm min} a=2.17$ and the $k a = 16 k_{\rm min} a=2.9$, as these values of the wave number probe the domains with predicted negative and positive quadratic responses. Figure \ref{response} displays the results of these perturbed simulations. The amplitude of the external potential energy was set to result in a 0.1$n_0$ amplitude of the linear part of the total density response. Panel (a) shows the resulting density distribution, with the average density, $n_0$, removed. The different shapes of the curves result from the fact that at the higher wave number the linear response dominates, while at the lower wave number the quadratic contribution is more significant. Removing the first-order perturbation to the density, the second-order contribution becomes visible in panel (b). Characteristic of the second-order response, we find oscillations with double frequency, compared to the perturbation. We observe that the second-order perturbation to the density changes sign between $k = 12 k_{\rm min}$ ($k a = 2.17$) and $k = 16 k_{\rm min}$ ($k a = 2.9$). The results clearly confirm the change of the sign of $\langle\tilde{n}(\textbf{r})\rangle^{(2)}$, in accordance with the change of sign of the 3-point structure function.

Finally we analyze the performance of the BHP model. As outlined in section 2, the BHP model offers an approximation for the triplet direct correlation function $c^{(3)}$, based on the ansatz (\ref{eq:bhp1}). In our computation of $c^{(3)}$ we followed the steps proposed in \cite{Hansen}, with the difference that instead of using a thermodynamically consistent variant of the HNC approach to compute  $\partial c^{(2)} / \partial n$, we rely on first-principles MD simulation  data. We first compute $S^{(2)}(k)$ in MD simulations with two different values of $\Gamma$, 800 $\pm$ 5 \%. From these static structure functions we calculate $c^{(2)}(k) = [1-1/S^{(2)}(k)] / n$ and then we execute an inverse Fourier-transformed that yields $c^{(2)}(r)$. The derivative of  $c^{(2)}(r)$ with respect to the density is obtained via $\partial c^{(2)} / \partial n = (\partial c^{(2)} / \partial \Gamma )(\partial \Gamma / \partial n)$. As the next step, we solve eq. (\ref{eq:bhp2}) for $t(r)$ using the steepest descent algorithm (also used in \cite{Hansen}), then compute its Fourier transform $t(k)$, and finally calculate $c^{(3)}({\bf k}_1,{\bf k}_2)$ as:
\begin{equation}
c^{(3)}({\bf k}_1,{\bf k}_2) = \frac{1}{(2 \pi)^3} \int t(k') t(|{\bf k}_1-{\bf k}'|) t(|{\bf k}_2+{\bf k}'|) {\rm d}{\bf k}',
\end{equation}
where ${\bf k}_1$ and ${\bf k}_2$ correspond to the actual (isoscele, or equilateral) geometry.

\begin{figure*}[floatfix,h!]
\begin{center}
\begin{tabular}{cc}
  \includegraphics[width=7.5cm]{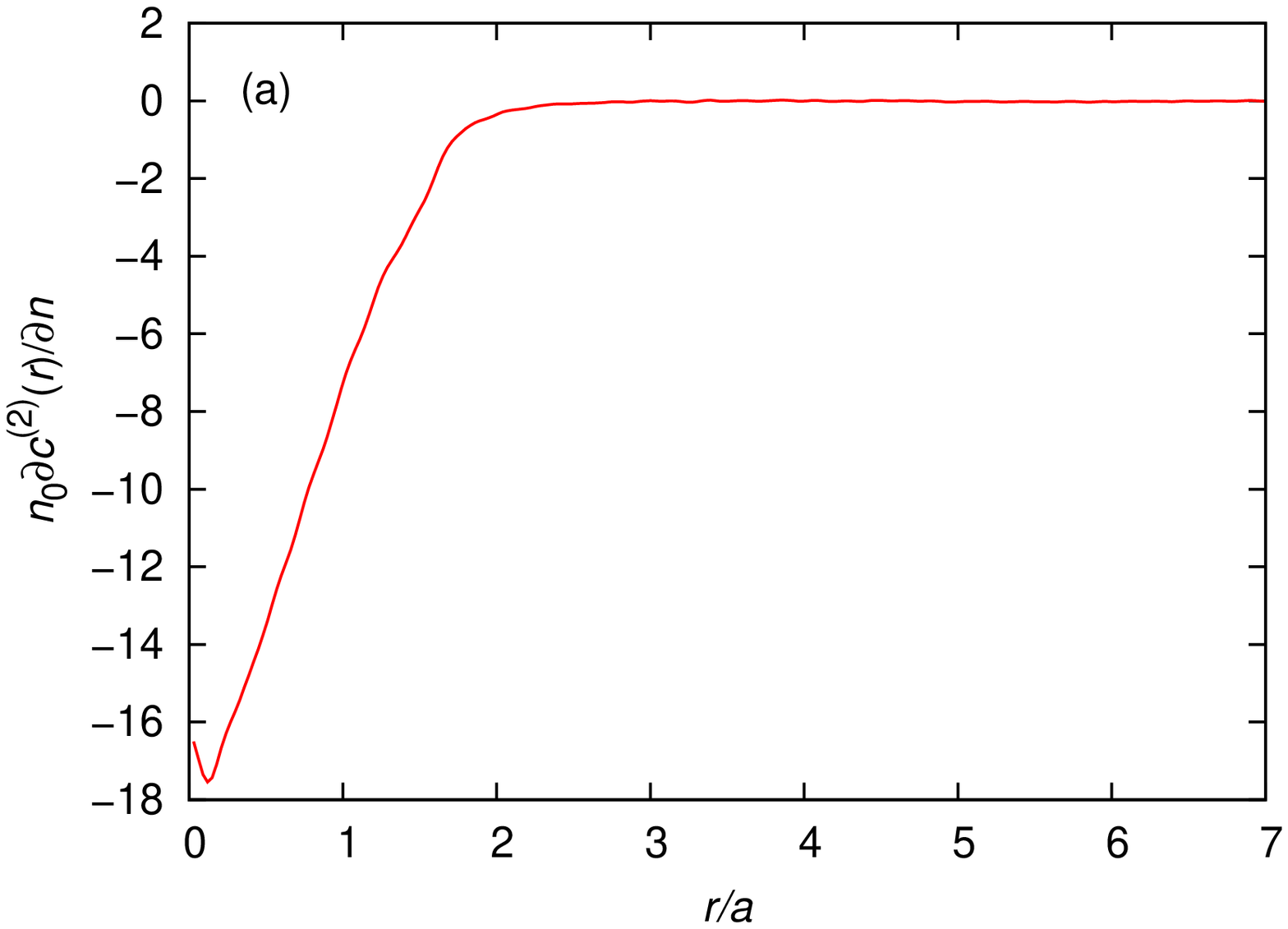}
 &
  \includegraphics[width=7.5cm]{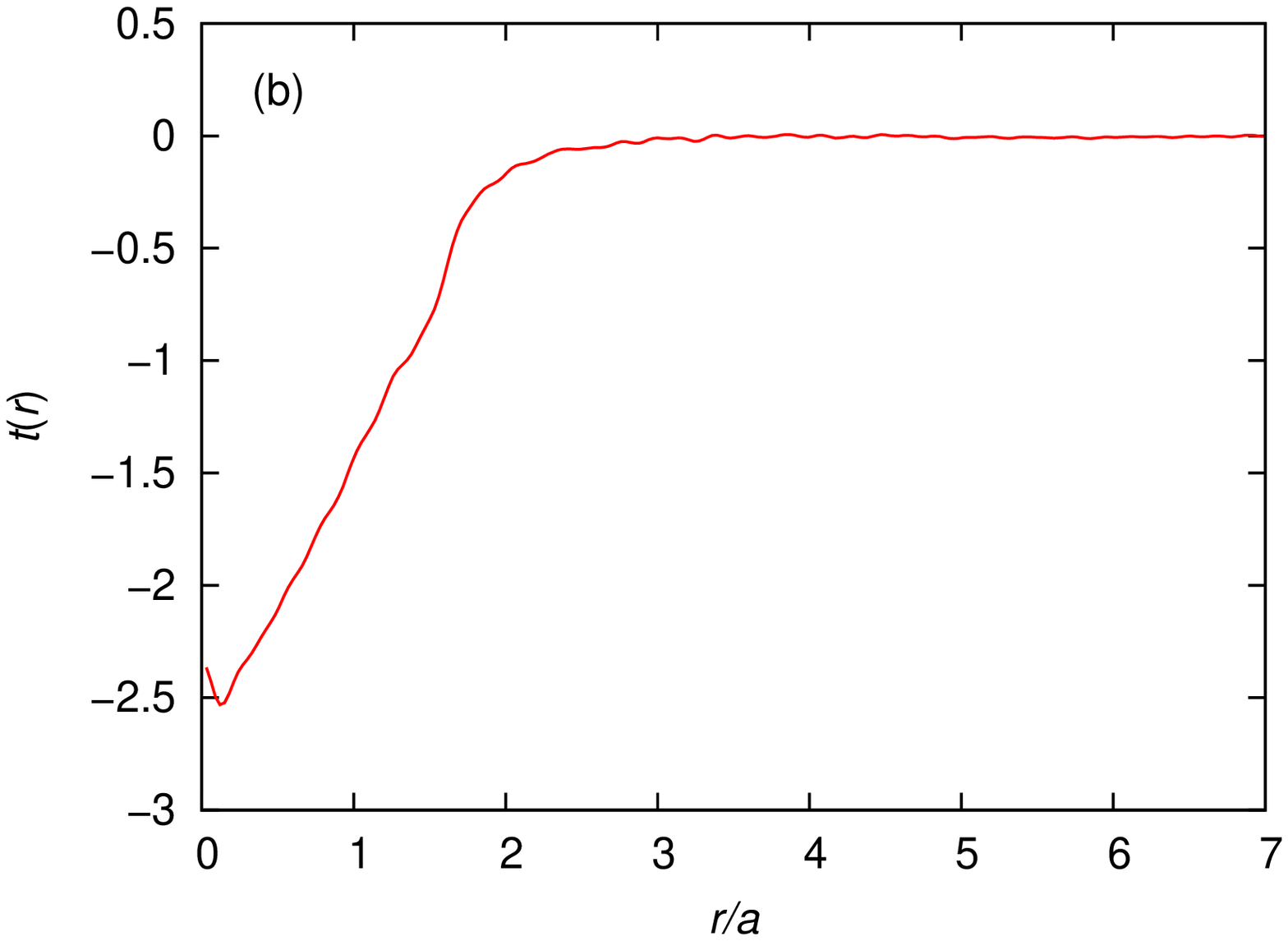}
 \\
\end{tabular}
\begin{tabular}{cc}
  \includegraphics[width=7.5cm]{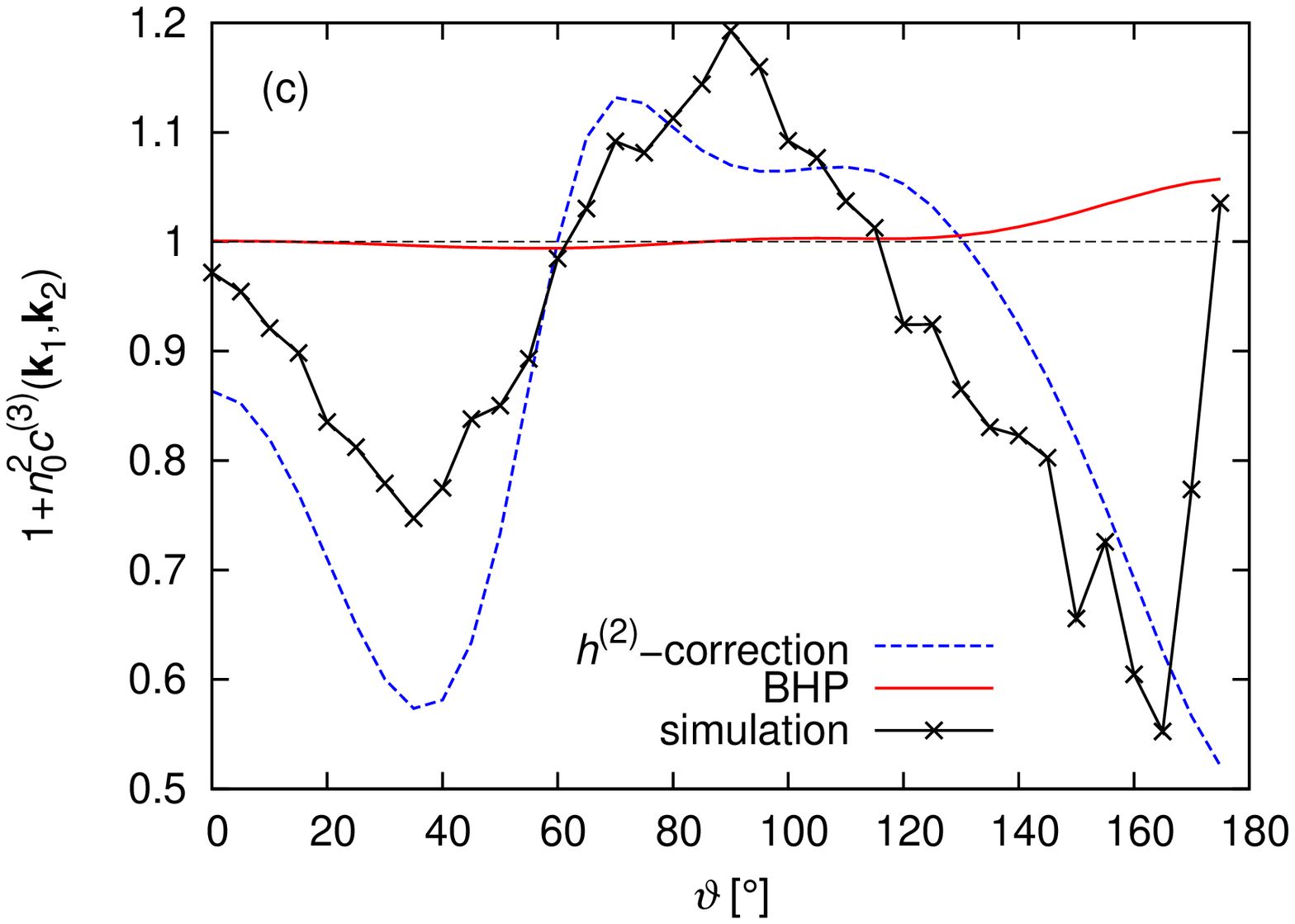}
 &
  \includegraphics[width=7.5cm]{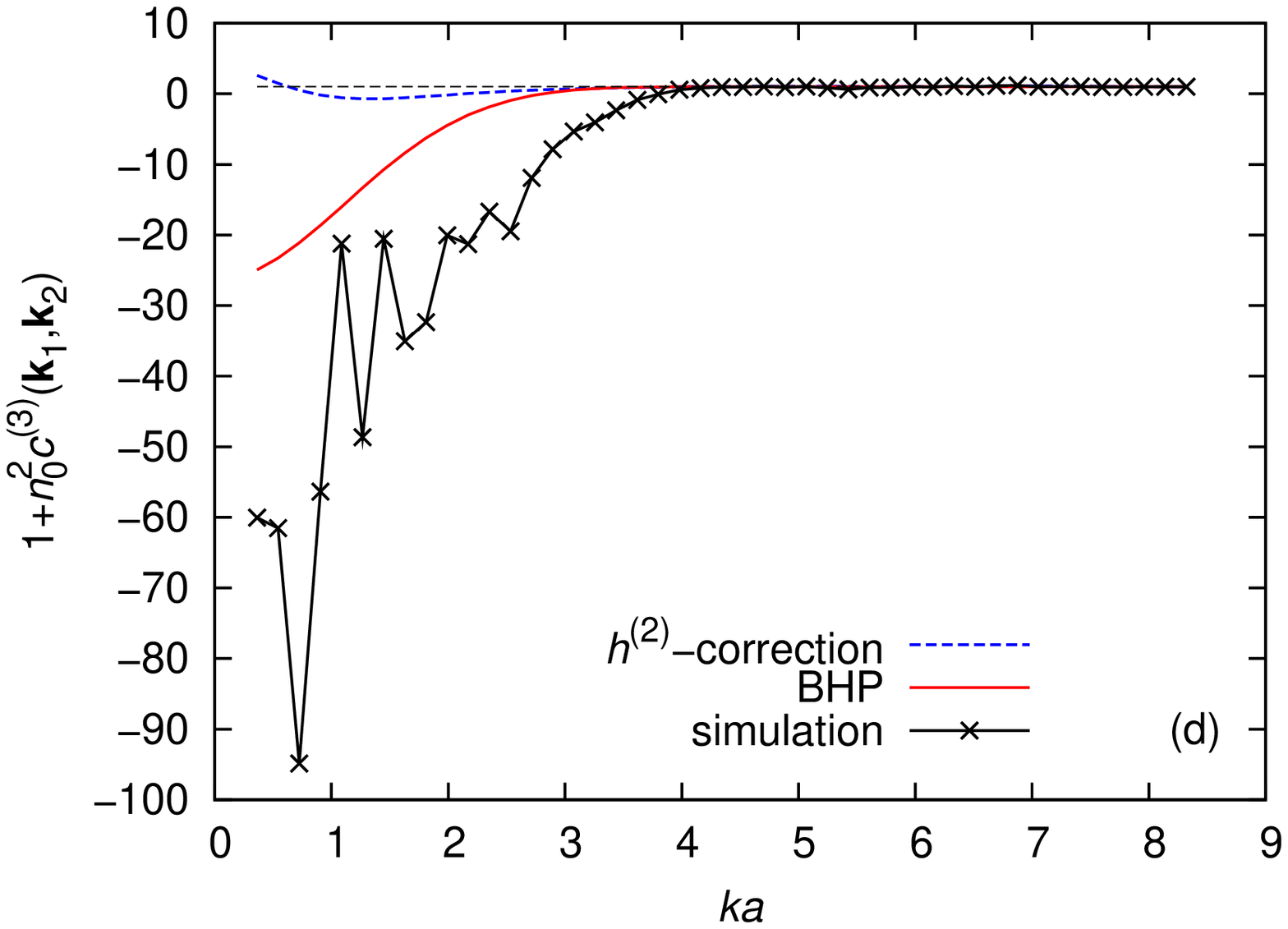}
 \\
\end{tabular}
\caption{(a) Normalized derivative of $c^{(2)}(r)$ with respect to density, (b) $t(r)$ obtained as the solution of eq. (\ref{eq:bhp2}), (c) and (d) $1 + n_0^2 c^{(3)}({\bf k}_1,{\bf k}_2)$ for wave vectors forming isoscele and equilateral triangles, respectively. In (c) $k^\ast a = 24 k_{\rm min}a = 4.34$. The ``$h^{(2)}$-correction'' is based on eq. (\ref{eq:hhh}).  $\Gamma = 800$ at $\kappa=3$.}
\label{fig:BHP}
\end{center}
\end{figure*}

The results of our calculations for the $\kappa=3$, $\Gamma=800$ case are presented in Fig.~\ref{fig:BHP}: panel (a) shows the normalized derivative $n_0 (\partial c^{(2)} / \partial n)$, panel (b) displays $t(r)$. The values of  $1 + n_0^2 c^{(3)}({\bf k}_1,{\bf k}_2)$ (recall that this equals to $S^{(3)} / S^{(2)} S^{(2)} S^{(2)}$), as obtained directly from the simulation, as computed from eq. (\ref{eq:hhh}) and as resulting from the BHP model are displayed in Figs.~\ref{fig:BHP}(c) and (d), respectively, for wave vectors forming isoscele and equilateral triangles.  In the case of the isoscele geometry, for which the data have been computed at $k^\ast a = 24 k_{\rm min}a = 4.34$, the ``$h^{(2)}$-correction'' gives a quite reasonable prediction for the ``true'' (direct simulation) values. On the other hand, the performance of the BHP model, for this specific case, is rather poor. In the case of the equilateral geometry we find an opposite behavior: the BHP performs reasonably, although it underestimates $1 + n_0^2 c^{(3)}({\bf k}_1,{\bf k}_2)$ by a factor of $\sim$3 as compared to the simulation results, while the ``$h^{(2)}$-correction'' is completely inaccurate.

\begin{figure*}[floatfix,h!]
\begin{center}
\begin{tabular}{cc}
  \includegraphics[width=7.5cm]{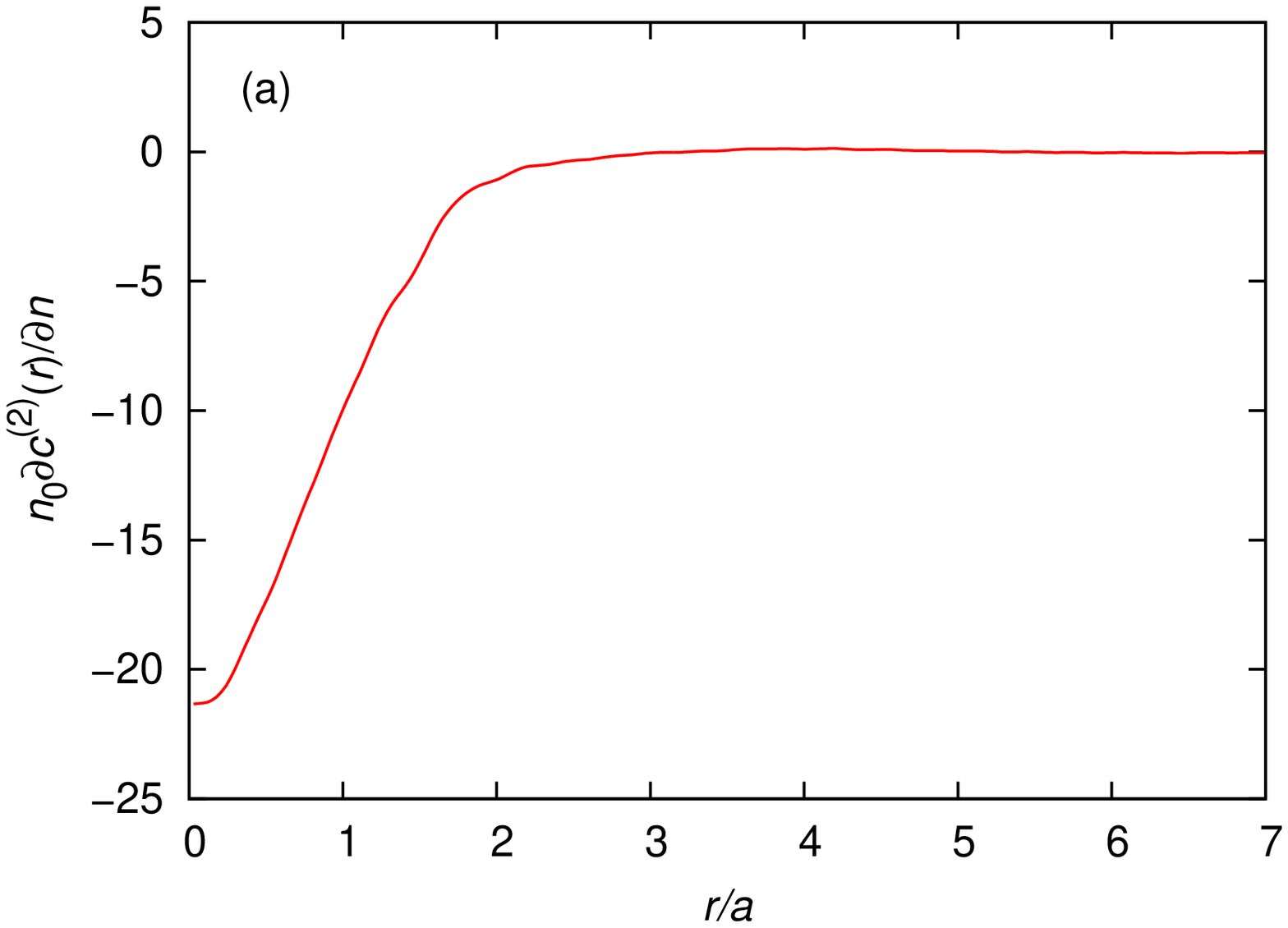}
 &
  \includegraphics[width=7.5cm]{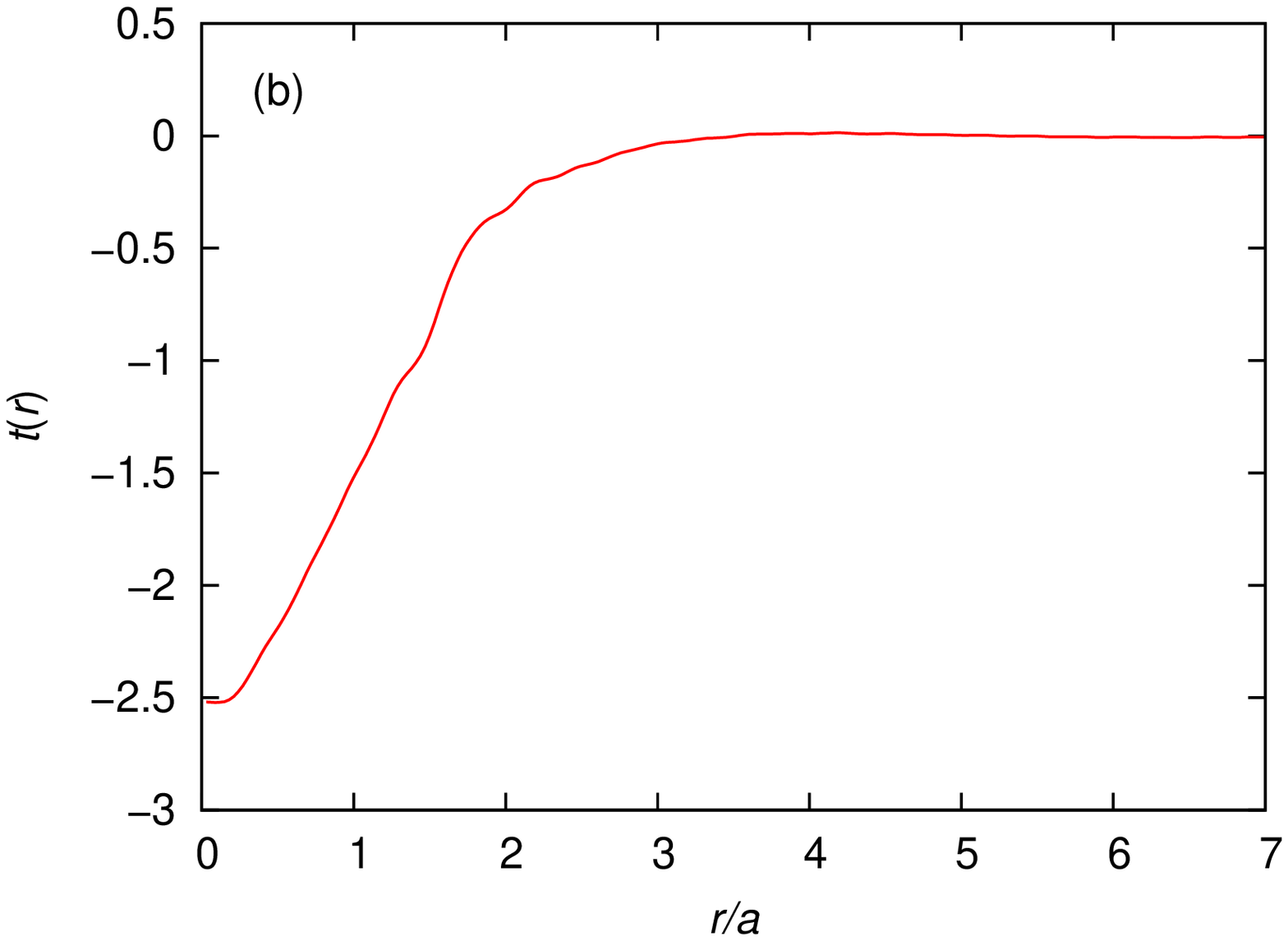}
 \\
\end{tabular}
\begin{tabular}{cc}
  \includegraphics[width=7.5cm]{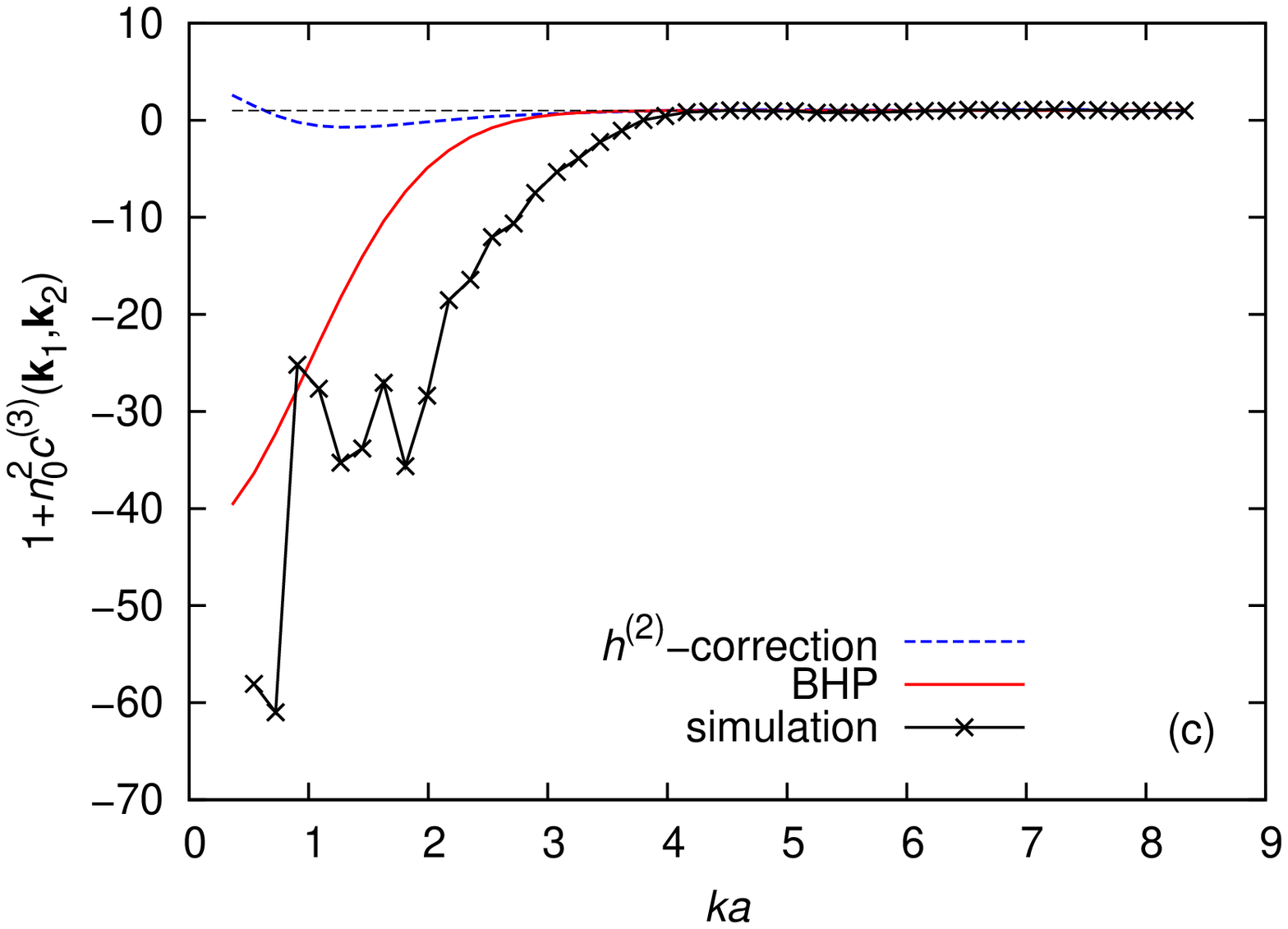}
\end{tabular}
\caption{(a) Normalized derivative of $c^{(2)}(r)$ with respect to density, (b) $t(r)$ obtained as the solution of eq. (\ref{eq:bhp2}), (c) $1 + n_0^2 c^{(3)}({\bf k}_1,{\bf k}_2)$ for wave vectors forming equilateral triangles. The ``$h^{(2)}$-correction'' is based on eq. (\ref{eq:hhh}). $\Gamma$=300, $\kappa=2$. }
\label{fig:BHP-k2}
\end{center}
\end{figure*}

As an additional case we also consider a system with $\kappa=2$ and $\Gamma=300$, with wave vectors forming equilateral triangles. The results of these computations are given in Fig.~\ref{fig:BHP-k2}: panel (a) shows the normalized derivative $n_0 (\partial c^{(2)} / \partial n)$, panel (b) displays $t(r)$, while panel (c) presents $1 + n_0^2 c^{(3)}({\bf k}_1,{\bf k}_2)$ as obtained from the simulations, as well as based on the ``$h^{(2)}$-correction'', and the BHP result. A somewhat better agreement between the simulation data and the BHP curve is obtained here, but the deviation between the results is still significant, i.e., the performance of the BHP model is not satisfactory for this case either.

In summary, the performance of the BHP model is rather weak for our system. To rule out the possibility that this behavior is a result of errors in our calculations, we have verified our procedures by comparing our results with those given in \cite{Hansen}, for a soft-sphere system. To be able to do so, we have developed an MD code for the $1/r^{12}$ potential and have carried out computations for the same parameters as in \cite{Hansen}. The results of this check are presented in the Appendix.

\section{Summary}

Using molecular dynamics simulations we have investigated the validity of the convolution approximation -- according to which in many-body systems the 3-point static structure function, $S^{(3)}(\textbf{k}_{1},\textbf{k}_{2})$, can approximately be factorized in terms of the 2-point counterpart, $S^{(2)}(\textbf{k}_{1})$. In our work we have also examined the factorization of the triplet correlation function $h^{(3)}(\textbf{k}_{1},\textbf{k}_{2})$.

As a model system we have adopted 3-dimensional strongly-coupled Yukawa liquids, the computations have been carried out for coupling parameter values $\Gamma = 80$ and 800 at $\kappa=3$, as well as for $\Gamma = 30$ and 300 at $\kappa=2$. We have tested the accuracy of the factorization for specific arrangements of the wave vectors $\textbf{k}_{1}$, $\textbf{k}_{2}$ and  $\textbf{k}_{0}$: (i) forming isosceles and (ii) equilateral triangles.

We have found a notable signature of  the breakdown of factorization: the appearance of negative values of $S^{(3)}(\textbf{k}_{1},\textbf{k}_{2})$, whereas the approximate factorized form is restricted to positive values. Based on the Fluctuation-Dissipation Theorem these negative values imply that the quadratic part of the density response of the system changes sign with wave number. Our simulations that incorporate an external potential energy perturbation clearly confirmed this behavior.

We have found that the BHP model \cite{Hansen} does not perform well for the Yukawa liquids with the specific parameters considered here. Taking into account the fact that the results of our computational procedures have been cross-checked with those obtained in \cite{Hansen} for the soft-sphere potential, the reason of the discrepancies could be sought for in the validity of the ansatz (\ref{eq:bhp1}) adopted in the BHP model.

\section{Acknowledgment}

This work was supported by the grants OTKA-K-105476, OTKA-K-115805, and NSF PHYS-1105005.

\section{Appendix}

\begin{figure*}[floatfix,h!]
\begin{center}
\begin{tabular}{cc}
  \includegraphics[width=7.5cm]{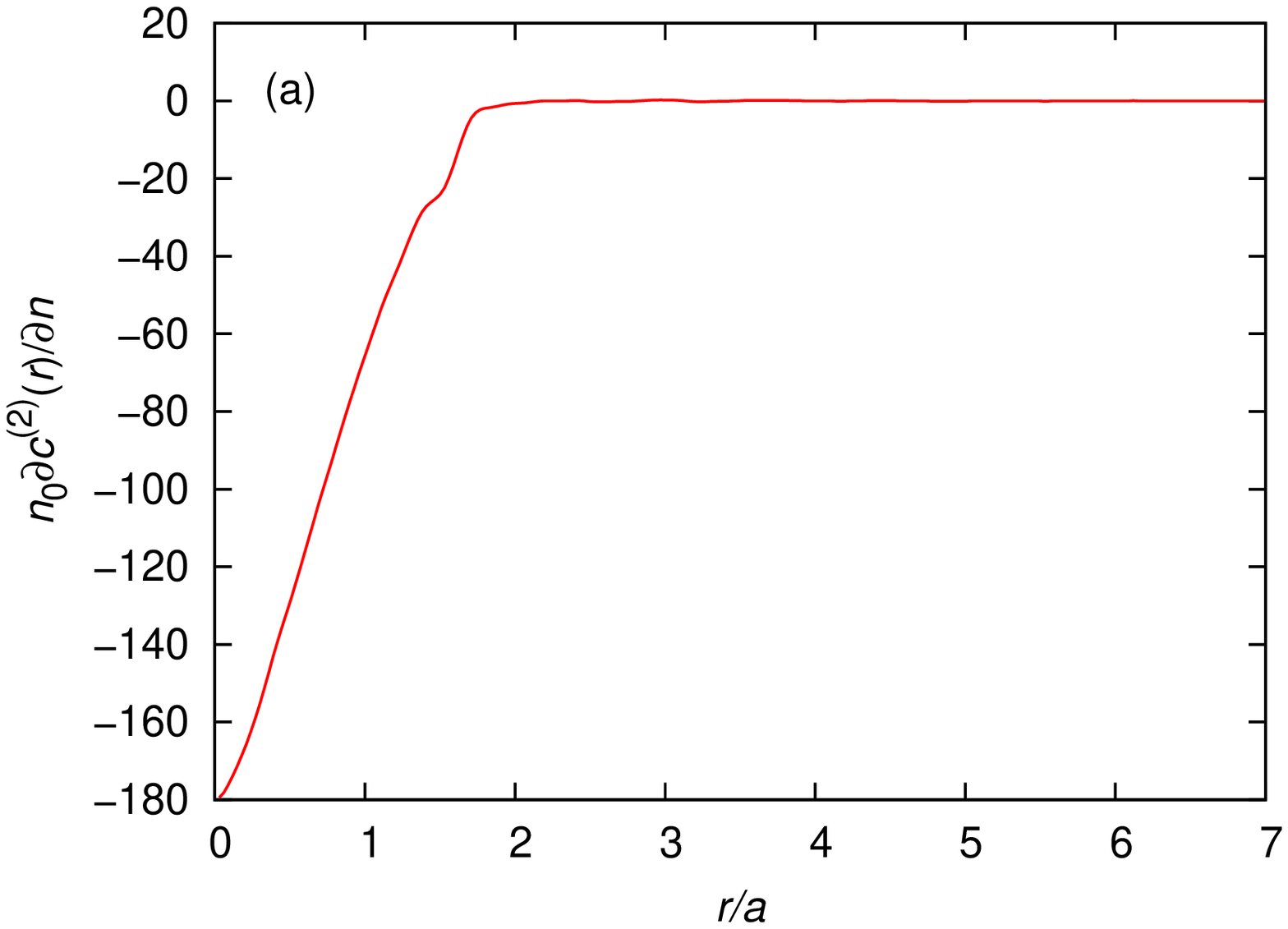}
 &
  \includegraphics[width=7.5cm]{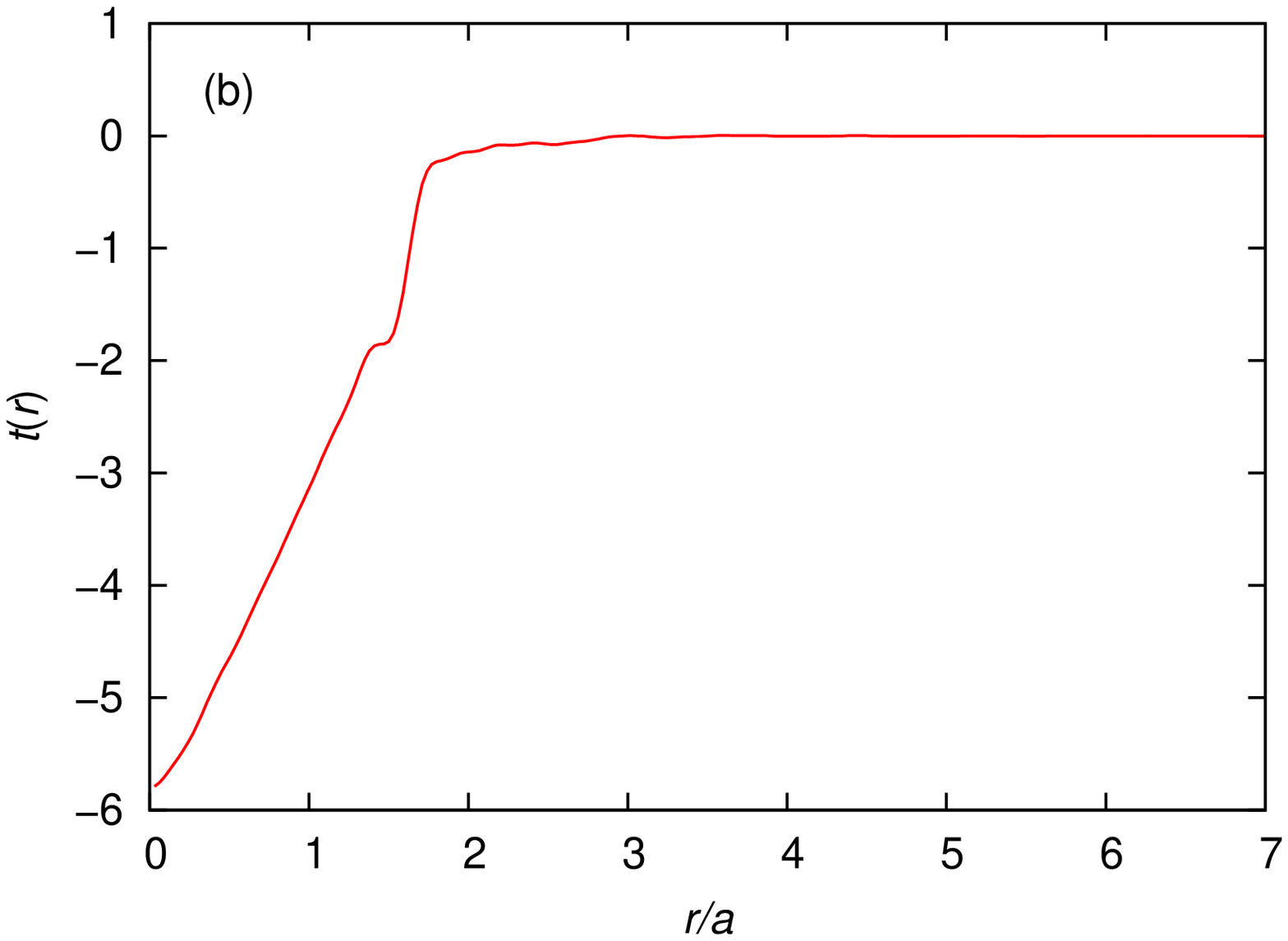}
 \\
\end{tabular}
\begin{tabular}{cc}
  \includegraphics[width=7.5cm]{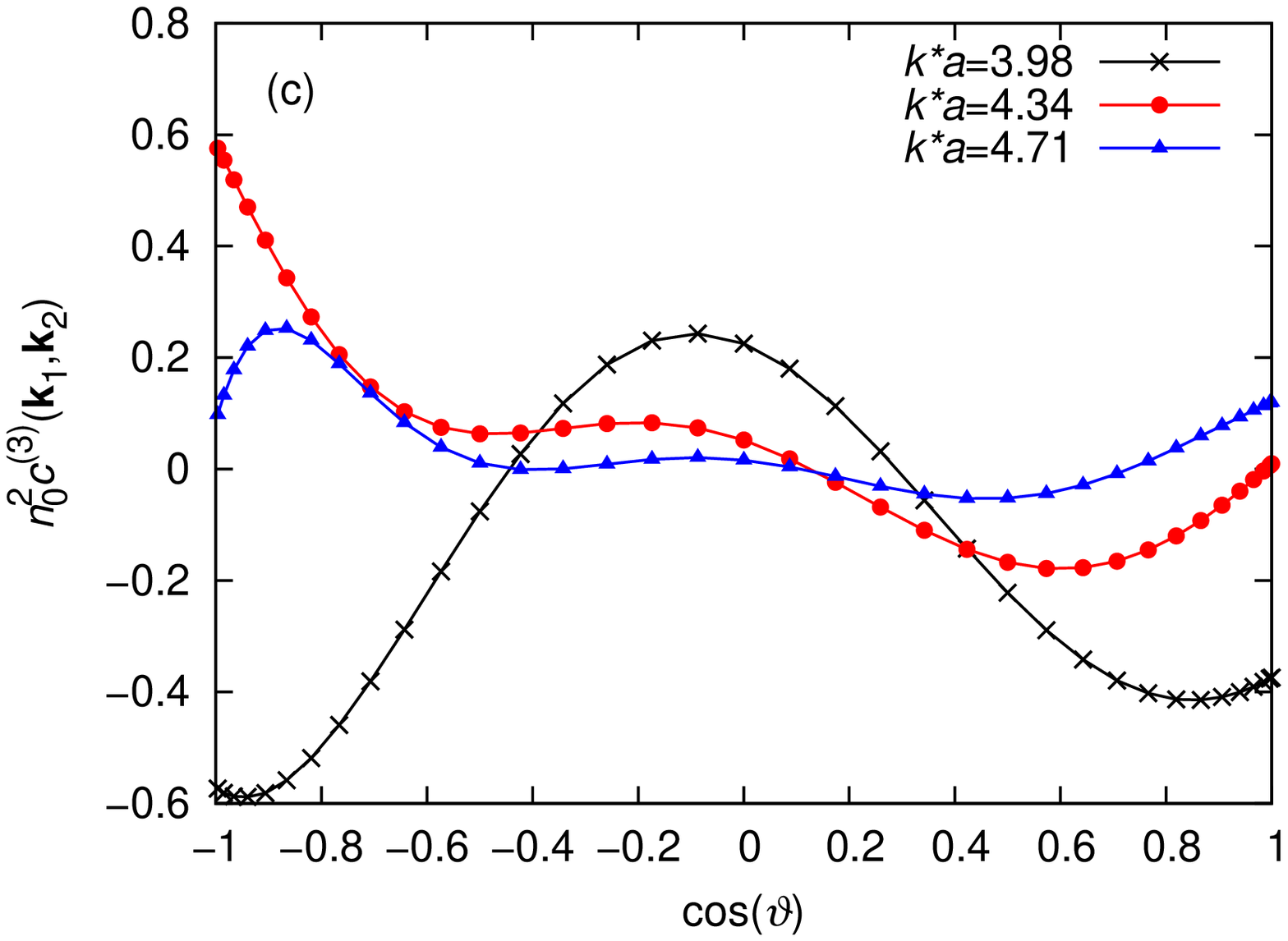}
 &
  \includegraphics[width=7.5cm]{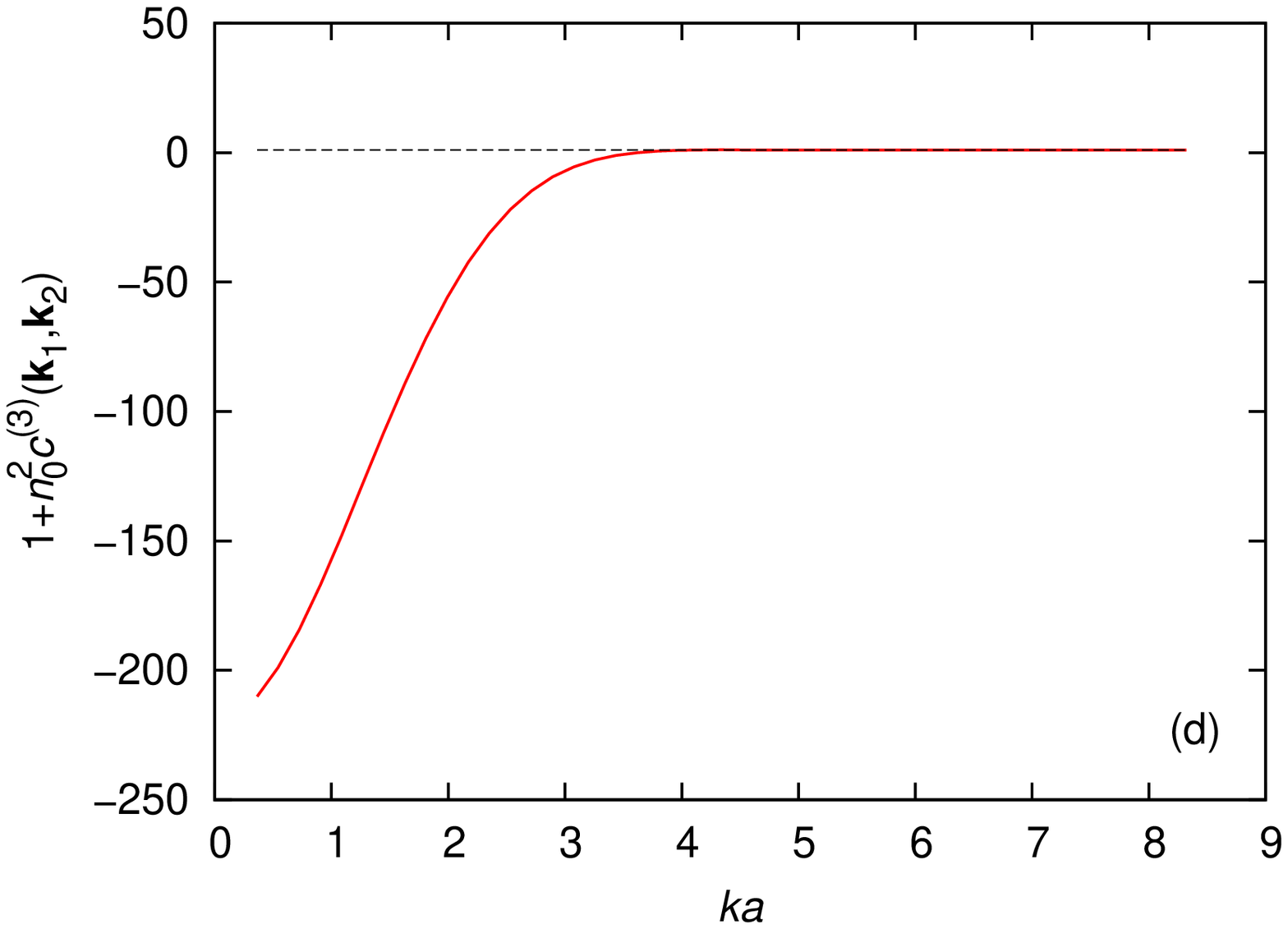}
 \\
\end{tabular}
\caption{Normalized derivative of $c^{(2)}(r)$ with respect to density, (b) $t(r)$ obtained as the solution of eq. (\ref{eq:bhp2}), (c) and (d) $1 + n_0^2 c^{(3)}({\bf k}_1,{\bf k}_2)$ for wave vectors forming isoscele and equilateral triangles, respectively. In (c) data for different values of the wave number $k^\ast a$ are displayed, and, in order to make comparison with the original data (figures 2 and 3 of  \cite{Hansen}) easier the data are presented here as a function of $\cos(\vartheta)$. Panel (a) is to be compared with figure 1 of \cite{Hansen} and panel (d) is to be compared with figure 4 of the same work.}
\label{fig:app}
\end{center}
\end{figure*}

As mentioned above, we have developed an MD simulation code, to be able to verify our computational procedures via cross-checking our results with those given in \cite{Hansen}, obtained for a soft-sphere potential. Our computations have been carried out for $\gamma=1.15$ (Barrat {\it et al.} \cite{Hansen} has considered values of 1.13, 1.15, and 1.17). Our results, shown in Fig.~\ref{fig:app} indicate a very good agreement with those presented in \cite{Hansen}: $t(r)$, shown in panel (b), is to be compared with figure 1 of  \cite{Hansen}, $n_0^2 c^{(3)}({\bf k}_1,{\bf k}_2)$, shown in panel (c), is to be compared with figures 2 and 3 of  \cite{Hansen}, which correspond, respectively, to slightly lower ($\gamma=1.13$) and slightly higher ($\gamma=1.17$) values of the control parameter of the system. Finally, panel (d) of Fig.~\ref{fig:app} is to be compared with figure 4 of \cite{Hansen}. We have found that for the isosceles case the shape of the $n_0^2 c^{(3)}$ vs. $\cos(\vartheta)$ curve is very sensitive on the wave number value, see Fig.~\ref{fig:app}(c). In our simulations the position of the first peak of the linear static structure function $S(k)$ was identified to be $k^\ast a = 4.34.$ In \cite{Hansen} it was also stated that the data correspond to the first peak of $S(k)$, however, the exact value was not given in \cite{Hansen}. Therefore here we compare the results obtained at different values of $k^\ast a$ in Fig.~\ref{fig:app}(c). Among the three curves the best agreement with the data of Barrat {\it et al.} \cite{Hansen} is found for $k^\ast a = 3.98$. The good reproducibility of the data presented in \cite{Hansen} verifies our computational techniques.

\end{document}